\documentclass[11pt]{article}

\usepackage[final]{acl}

\usepackage{times}
\usepackage{latexsym}
\usepackage[T1]{fontenc}
\usepackage[utf8]{inputenc}
\usepackage{microtype}
\usepackage{inconsolata}
\usepackage{graphicx}
\usepackage{multirow}
\usepackage{booktabs}
\usepackage{adjustbox}
\usepackage{amsmath}
\usepackage{amssymb}
\usepackage{xcolor}
\usepackage{colortbl}
\usepackage{rotating}
\usepackage{cuted}
\usepackage{tcolorbox}
\usepackage{subcaption}
\usepackage[titletoc,page]{appendix}
\DeclareMathOperator*{\argmax}{arg\,max}

\definecolor{goldrow}{RGB}{255,248,220}
\definecolor{pooledrow}{RGB}{230,240,255}
\definecolor{highlight}{RGB}{255,235,205}
\newcommand{\gold}[1]{\cellcolor{goldrow}#1}
\newcommand{\pooled}[1]{\cellcolor{pooledrow}#1}
\newcommand{\hgold}[1]{\cellcolor{goldrow}#1}
\newcommand{\hpooled}[1]{\cellcolor{pooledrow}#1}
\usepackage{xspace} 
\newcommand{\sys}{\textsc{Obliq-IR}\xspace}
\newcommand{\xr}[1]{#1}

\title{OBLIQ-IR: Training a Dense Retriever for Oblique Queries}

\author{
  \textbf{Mahmoud Abdalla\textsuperscript{1}\thanks{Mahmoud Abdalla and Abdelrahman Abdallah contributed equally.},
  Abdelrahman Abdallah\textsuperscript{2}\footnotemark[1],
  Shaimaa Sedek\textsuperscript{3},
  Adam Jatowt\textsuperscript{2}} \\
  \textsuperscript{1}Chungbuk National University \quad
  \textsuperscript{2}University of Innsbruck \quad
  \textsuperscript{3}Independent Researcher \\
  \texttt{\{abdelrahman.abdallah,adam.jatowt\}@uibk.ac.at}
}

\begin{document}
\maketitle
\begin{strip}
  \centering
  \includegraphics[width=0.8\textwidth]{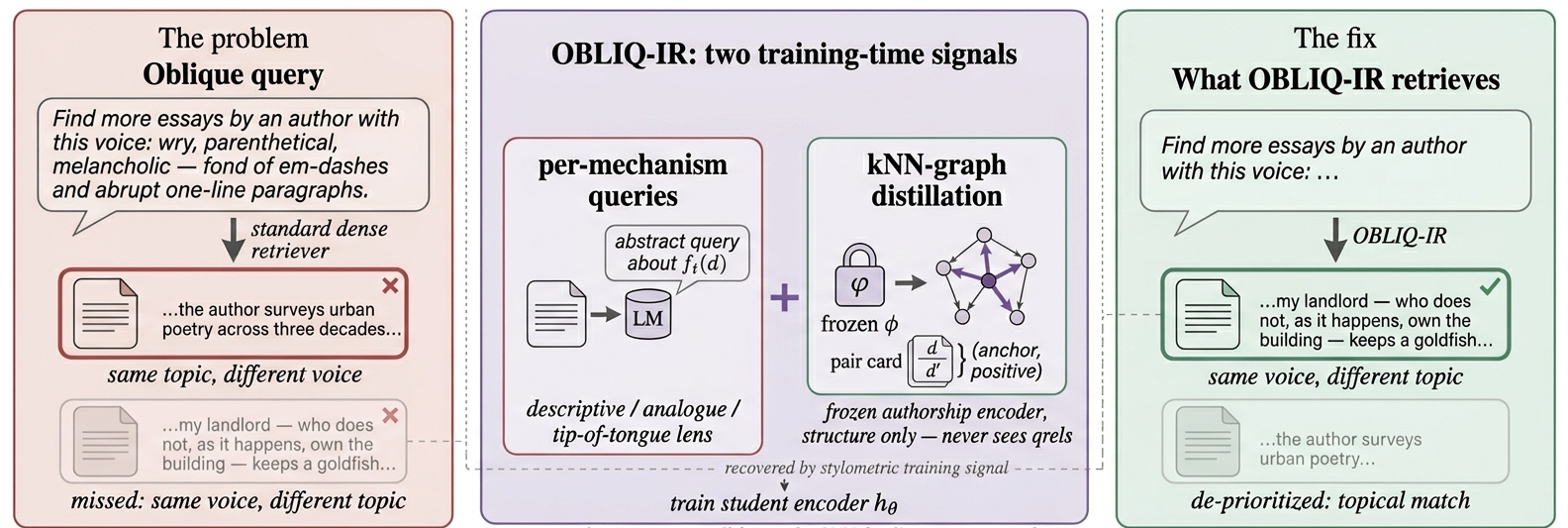}
  \captionof{figure}{Oblique queries are defined by a latent attribute the document barely surfaces. \textbf{Left:} a standard dense retriever, asked for documents matching an author's voice, returns a topical match and buries the true stylistic match. \textbf{Center:} OBLIQ-IR adds two training-time signals: per-mechanism synthetic queries written through one of three lenses (descriptive, analogue, tip-of-the-tongue) and kNN-graph distillation from a frozen authorship encoder $\phi$, whose nearest-neighbor structure becomes anchor-positive training pairs for the student encoder $h_\theta$. \textit{Per-mechanism queries cover all four tasks; kNN distillation is writing only.} \textbf{Right:} the same query against OBLIQ-IR promotes the stylistic match to the top and de-prioritizes the topical one.}
  \label{fig:teaser}
\end{strip}
\begin{abstract}
Oblique retrieval, as exemplified by OBLIQ-Bench, asks a retriever to find documents whose relevance is determined by a latent attribute (an implicit stance, an analogous reasoning technique, an authorial fingerprint, or a vague tip-of-the-tongue recollection) that has little or no surface expression in the document. State-of-the-art dense encoders and agentic search pipelines built around frontier language models exhibit a large first-stage bottleneck on these tasks, while the same language models reliably verify relevance when shown candidates. We address this with OBLIQ-IR, a single-vector dense retriever whose training mixture combines per-mechanism synthetic queries with a new form of cross-model supervision: kNN-graph distillation from a frozen authorship encoder, which transfers a style-versus-topic inductive bias into the student. A 3B retriever fine-tuned reaches 0.211 NDCG@10 on Writing-Style, 0.171 on Math, 0.177 on Twitter, and 0.281 on Congress, improving over the GPT-5.2 Multi-Hop Agent by \xr{0.010 to 0.150} NDCG@10 and over Gemini-2-Embedding by 0.027 to 0.222 NDCG@10 on every reported task. 
\footnote{The code, data and checkpoints are available \url{https://github.com/DataScienceUIBK/obliq-ir} .}
\end{abstract}

\section{Introduction}
\label{sec:introduction}

Information retrieval has long focused on topical match: returning documents whose content overlaps with the query. This paradigm however breaks for so-called oblique queries \citep{oblique2026} or inferential queries \citep{inferentialQA}. The relevance of such queries is determined by a latent property (e.g., an implicit stance, an abstract proof strategy, an authorial style, a tip-of-the-tongue rhetorical dynamic, or required inference) that has little or no surface expression in a relevant document (Figure~\ref{fig:teaser}).

OBLIQ-Bench \citep{oblique2026}, with an accompanying suite of five tasks over long-tail corpora, reports a large gap between retrieval and verification: pipelines built around frontier language models score close to zero NDCG@10 on most tasks, while the same language models reliably recognize relevance once a candidate is presented. The bottleneck is first-stage search; if the relevant document is not in the candidate pool, no amount of downstream reasoning can recover it. \citet{oblique2026} call for ``architectures that make latent document attributes available at search time.''

In this work, we present \sys, the first bi-encoder retriever trained specifically for oblique queries. The key element is a per-mechanism training mixture (\S\ref{sec:method}) that combines two complementary signals: (1) \emph{per-mechanism synthetic queries}, generated by prompting an open instruction-tuned language model through one of three lenses---descriptive, analogue, or tip-of-the-tongue---that match the OBLIQ query mechanisms, so that each training query exercises the latent attribute of the target task rather than its surface vocabulary; and (2) \emph{kNN-graph distillation from a frozen authorship encoder}, applied to the writing-style task where no topical signal exists. Because an authorship encoder contrastively trained on a large authorship corpus already separates style from topic in its representation space, we use only the topology of its kNN graph over the writing corpus---not its similarity values---to construct anchor-positive pairs for the student. The teacher's role is purely structural; its scores are discarded.

To our knowledge, no prior work in dense retrieval has used a frozen authorship encoder as a training-time structural signal in this way. We train \sys by fine-tuning a 3B bi-encoder \citep{nvembed} on a mixture of public retrieval data and our per-mechanism synthetic queries, with kNN-graph distillation applied to the writing task only. \xr{We evaluate all five OBLIQ-Bench tasks} (\S\ref{sec:experiments})\xr{; \sys improves on four of them, and our WildChat-Errors results are reported in Appendix~\ref{app:wildchat}}. \sys achieves a new best non-oracle NDCG@10 on every reported task, reaching 0.211 on Writing-Style, 0.171 on Math, 0.177 on Twitter, and 0.281 on Congress, improving over the GPT-5.2 Multi-Hop Agent by \xr{0.010 to 0.150} NDCG@10 and over Gemini-2-Embedding by 0.027 to 0.222 NDCG@10. Paired bootstrap intervals confirm the gains as significant on Writing and Congress and marginal on Math and Twitter. \xr{All results are obtained in a supervised setting: the mechanism lenses, instruction prefixes, and reranker policy are chosen with knowledge of the benchmark's mechanism taxonomy, so we make no claim of zero-shot transfer to unseen oblique tasks.} Our proposed method can be readily adapted to other latent-attribute retrieval tasks where a small frozen specialist captures the target signal\xr{; \S\ref{sec:when-distill} gives a cheap diagnostic for deciding when that is the case}.

\section{Problem Formulation}
\label{sec:formulation}

\paragraph{Retrieval task.}
We follow the OBLIQ-Bench formalization \citep{oblique2026}. A corpus $\mathcal{C}=\{d_1,\ldots,d_N\}$ is given. At query time a retriever receives a query $q$ and an integer $k\ll N$ and returns an ordered list of $k$ documents. Quality is measured by NDCG@$k$ and Recall@$k$ under both Gold and Pooled relevance judgments \citep{oblique2026}.

\paragraph{Oblique queries.}
A query $q$ is \emph{oblique} when the property that determines whether a candidate document $d$ is relevant is a latent function $f_t(d)$ that has little surface expression in $d$. OBLIQ-Bench enumerates four such latent properties that we cover in this paper: implicit stance toward a geopolitical conflict (Twitter-Conflict), the abstract reasoning technique used to solve a math problem (Math Meta-Program), the authorial style of a prose passage (Writing-Style), and the rhetorical dynamic of a Congressional exchange given a lossy recollection of it (Congress Hearings). The benchmark groups these tasks by query \emph{mechanism}: descriptive queries (Twitter), analogue queries (Math, Writing), and tip-of-the-tongue queries (Congress).

\paragraph{Single-vector retriever.}
We learn an encoder $h_\theta:\mathrm{text}\to\mathbb{R}^d$ shared across queries and documents and rank by cosine similarity $s_\theta(q,d)=\cos(h_\theta(q),h_\theta(d))$. The training objective is the contrastive log-likelihood used by recent reasoning-oriented retrievers \citep{reasonir}:
\begin{equation}
\ell(q) \;=\; -\log \frac{\exp\bigl(\tau\, s_\theta(q,d^+)\bigr)}{\sum_{d\in\{d^+\}\cup D^-}\exp\bigl(\tau\, s_\theta(q,d)\bigr)},
\label{eq:infonce}
\end{equation}
where $\tau$ is a temperature scale, $d^+$ is a mined positive document, and $D^-$ is a set of mined and in-batch negatives; the specific values appear in the Experiment Setup.
\begin{figure*}[t]
\centering
\includegraphics[width=0.8\textwidth]{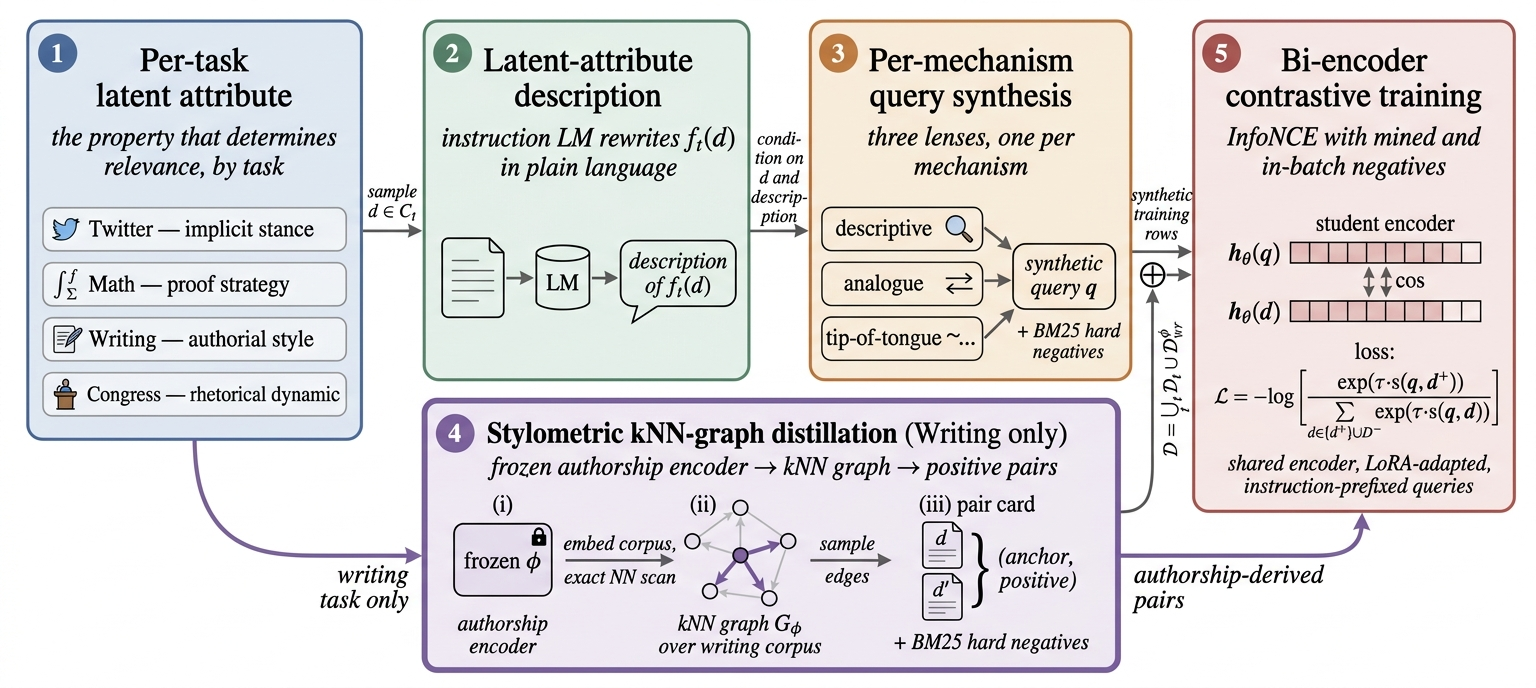}

\caption{\textbf{OBLIQ-IR training pipeline.} The five stages: (1) identify the per-task latent attribute; (2) write a one-paragraph description of that attribute for each sampled document with an instruction language model; (3) draft a query through one of three mechanism lenses (descriptive, analogue, or tip-of-the-tongue); (4) for the writing task only, also build training pairs from the kNN graph of a frozen authorship encoder; (5) train a bi-encoder with the contrastive objective. The inset on the right shows that writing NDCG@10 rises from 0.096 with no distillation to 0.211 at the dose used by OBLIQ-IR, and that larger doses overfit on writing while degrading the other tasks.}
\label{fig:method}
\end{figure*}
\section{Method}
\label{sec:method}

OBLIQ-IR is trained on a mixture of two complementary signals: per-mechanism synthetic queries (\S\ref{sec:permech}) and cross-model supervision distilled from a frozen authorship encoder through its kNN graph (\S\ref{sec:luar}). Section~\ref{sec:obj} specifies the contrastive training objective applied to the combined mixture. Figure~\ref{fig:method} summarizes the full pipeline.

\subsection{Per-mechanism synthetic queries}
\label{sec:permech}
For every task $t$ we want a training set whose queries are answerable only by reasoning about the latent attribute $f_t$, not by surface overlap with the document. We obtain such queries with a two-stage instruction-language-model pipeline. \textbf{Stage~A} reads a document $d\in\mathcal{C}_t$ and writes a short abstract description of $f_t(d)$. \textbf{Stage~B} conditions on $d$ and on that description and writes a query $q$ that would be satisfied by any other document with the same latent attribute, with an explicit instruction to avoid named entities and rare vocabulary from $d$. The Stage~B prompt is one of three lenses, matched to the query mechanism reported by OBLIQ-Bench \citep{oblique2026}: \textbf{(1) } the \emph{descriptive} lens prompts for an abstract paraphrase of an implicit property of the document (used for Twitter, where the property is implicit stance toward a conflict);
\textbf{(2) }the \emph{analogue} lens prompts for a query that requests other documents sharing the abstract structure of $d$, regardless of surface topic (used for Math, where the structure is the proof strategy, and Writing, where it is the authorial style);
\textbf{(3) }the \emph{tip-of-the-tongue} lens prompts for a conversational, lossy recollection of the rhetorical dynamic of $d$ that omits names, dates, and verbatim phrases (used for Congress).

\xr{We verify that the instruction is actually followed rather than merely issued: measured against their source documents, the synthetic queries leak entities in 1.7--6.4\% of cases and their lexical overlap is comparable to that of the benchmark's own oblique queries on three of the four tasks (Appendix~\ref{app:query-quality}).} For each $(q,d)$ pair we mine hard negatives by BM25 over $\mathcal{C}_t$ to give the contrastive objective a useful gradient. The training set for task $t$ is $\mathcal{D}_t=\{(q,d,D^-)\}$; the per-mechanism lens is what makes $\mathcal{D}_t$ exercise $f_t$ rather than topical match.

\subsection{Stylometric knowledge distillation}
\label{sec:luar}
The writing task is the only one of the four whose latent attribute is purely stylistic rather than semantic. Topical descriptive or analogue lenses cannot expose authorship: the same author writes about different topics, and a query that describes one of those topics will recall topically related but stylistically unrelated documents. We address this with cross-model supervision from a frozen authorship encoder $\phi$ pretrained on a large authorship corpus \citep{luarmud}. The encoder is contrastively trained so that two passages by the same author across topics are closer under $\phi$ than two passages on the same topic by different authors; it therefore captures the style-versus-topic axis we need.

\paragraph{Style-space kNN graph.}
We use $\phi$ to define a directed similarity graph over the writing corpus and then distill that graph into the student. Let $\mathcal{C}_{\mathrm{wr}}$ be the writing corpus and let $\phi:\mathrm{text}\to\mathbb{R}^{d_\phi}$ be the frozen authorship encoder. We construct the style-space kNN graph
\begin{equation}
G_\phi \;=\; \bigl(\mathcal{C}_{\mathrm{wr}},\; E_\phi\bigr),
\label{eq:graph-def}
\end{equation}
whose edges are the directed nearest-neighbor relations under cosine similarity of $\phi$ embeddings:
\begin{equation}
E_\phi \;=\; \bigl\{(d,d')\;:\;d'\in\mathrm{NN}_\phi^{\,k}(d)\bigr\},
\label{eq:graph-edges}
\end{equation}
where
\begin{equation}
\mathrm{NN}_\phi^{\,k}(d) \;=\; \argmax_{S\subseteq \mathcal{C}_{\mathrm{wr}}\setminus\{d\},\,|S|=k}\;\sum_{d'\in S}\cos\bigl(\phi(d),\phi(d')\bigr)
\label{eq:nn-def}
\end{equation}
returns the $k$ documents most similar to $d$ in style space. The graph is sparse with out-degree $k$, and $k$ is a hyperparameter we set in the Experiment Setup. We pre-compute $\phi(d)$ for every $d\in\mathcal{C}_{\mathrm{wr}}$ once, build $G_\phi$ with an exact dot-product scan because $|\mathcal{C}_{\mathrm{wr}}|$ is small enough to permit it, and then discard $\phi$ for the rest of training.

\paragraph{From graph to training pairs.}
We convert the edges of $G_\phi$ into positive pairs for the student. Each directed edge $(d,d')\in E_\phi$ is interpreted as a stylistic anchor-positive pair: the student should rank $d'$ above unrelated documents when its query is the natural-language description of $d$'s style. We assemble the writing distillation subset by sampling a fixed number of edges uniformly at random:
\begin{equation}
\mathcal{D}_{\mathrm{wr}}^{\,\phi}\;\sim\;\mathrm{Uniform}(E_\phi),
\label{eq:sample-edges}
\end{equation}
and then expanding each sampled edge into a ReasonIR-format \citep{reasonir} training row by attaching mined BM25 hard negatives from $\mathcal{C}_{\mathrm{wr}}$:
\begin{equation}
(d,d')\;\longmapsto\;\bigl(q_d,\; d',\; D^-_{\mathrm{BM25}}(d')\bigr),
\label{eq:row-construction}
\end{equation}
where $q_d$ is a one-line natural-language description of $d$'s stylistic fingerprint produced by the same instruction language model used in Stage~A of Section~\ref{sec:permech}.

\paragraph{Why graph and not scores.}
The frozen authorship encoder is never asked about a query at training time and never reads the OBLIQ qrels; only the structure of $G_\phi$ is used. We refer to this as \emph{kNN-graph distillation} from a frozen specialist, to distinguish it from the score-distillation tradition of MarginMSE \citep{marginmse} and TAS-B \citep{hofstatter2021tas}, in which the teacher's pairwise margins are passed to the student as soft labels. The graph carries the inductive bias of $\phi$ (same-author across topics are close, same-topic across authors are far) without committing the student to the specialist's exact metric. A student that minimizes the contrastive objective (Eq.~\ref{eq:infonce}) on $\mathcal{D}_{\mathrm{wr}}^{\,\phi}$ is forced to allocate capacity to features that distinguish authors rather than features that distinguish topics, inheriting the specialist's inductive bias while remaining a single-vector retriever that can answer natural-language queries.

The combined training set is
\begin{equation}
\mathcal{D}\;=\;\Bigl(\bigcup_t \mathcal{D}_t\Bigr) \;\cup\; \mathcal{D}_{\mathrm{wr}}^{\,\phi}.
\label{eq:totalmix}
\end{equation}
The relative size of $\mathcal{D}_{\mathrm{wr}}^{\,\phi}$ matters: too small and the student does not see enough style examples; too large and the encoder overfits writing at the expense of the other tasks. We treat the dose as a hyperparameter; the sweep appears in Figure~\ref{fig:method} (right) and Section~\ref{sec:experiments}.

\subsection{Training objective}
\label{sec:obj}
We train a single bi-encoder $h_\theta$ on $\mathcal{D}$ with the contrastive log-likelihood of Eq.~\ref{eq:infonce}, with both mined and in-batch negatives. The same encoder is used for queries and documents and the same parameters are used for all four tasks. Task identity is supplied implicitly through a per-task instruction prefix prepended to the query at both training and inference time. This two-element query format, in which the encoder is conditioned on a short natural-language instruction that names the retrieval criterion, was introduced by INSTRUCTOR \citep{instructor} and adopted by ReasonIR \citep{reasonir}; the synthetic doc-to-query training pipeline that we extend with per-mechanism lenses follows the few-shot recipe of Promptagator \citep{promptagator}.

\subsection{Mechanism-selective reranking}
\label{sec:rerank}
At inference time the dense retriever returns the top-100 candidates per query, and a task-specific reranker $\pi_t$ reorders that pool. We consider three policies: $\mathrm{id}$ (identity, no reranking); $\mathrm{TR}$, a tournament-style listwise reranker \citep{tourrank} that asks an instruction language model to rank candidates by ``relevance''; and $\mathrm{LW}_t$, a single-pass task-aware listwise reranker whose system prompt names the latent attribute $f_t$ explicitly. The choice of $\pi_t$ for each task is made on a held-out development split by maximising development NDCG@10:
\begin{equation}
\pi_t^\ast \;=\; \argmax_{\pi\in\{\mathrm{id},\mathrm{TR},\mathrm{LW}_t\}} \; \mathrm{NDCG@10}\bigl(\pi \circ h_\theta;\,\mathrm{dev}_t\bigr).
\label{eq:policy}
\end{equation}
We refer to the full paper-final pipeline (OBLIQ-IR dense retriever followed by $\pi_t^\ast$) as \textbf{OBLIQ-IR (full)}. The selected policies are $\pi_t^\ast = \mathrm{TR}$ for Twitter, Math, and Congress, and $\pi_t^\ast = \mathrm{id}$ for Writing; the reranker is the same 27B open-weights language model used in Stage~B of synthetic data generation. The motivating hypothesis, that LLM rerankers operate on a notion of relevance that is itself topical and therefore help tasks whose latent attribute is approximately topical and hurt tasks whose latent attribute is orthogonal to topic, is confirmed quantitatively in Section~\ref{sec:analysis} and Appendix~\ref{app:rerank-cost-quality}.

\begin{table*}[t]
\centering
\small
\caption{Twitter-Conflict (descriptive queries). Each metric shows \colorbox{goldrow}{Gold (G)} / \colorbox{pooledrow}{Pooled (P)} evaluation. \colorbox{highlight}{Key metrics} highlighted. \textbf{Bold} = best non-oracle row. Baseline numbers reproduced from \citet{oblique2026}.}
\label{tab:twitter-results}
\begin{adjustbox}{max width=0.85\textwidth}
\begin{tabular}{l cc cc ccc}
\toprule
& \multicolumn{2}{c}{\cellcolor{highlight}\textbf{NDCG@10}} & \multicolumn{2}{c}{\textbf{NDCG@50}} & \multicolumn{3}{c}{\textbf{Recall}} \\
\cmidrule(lr){2-3} \cmidrule(lr){4-5} \cmidrule(lr){6-8}
\textbf{Pipeline} & \hgold{\textbf{G}} & \hpooled{\textbf{P}} & \gold{G} & \pooled{P} & \cellcolor{highlight}\textbf{@10} & @50 & @100 \\
\midrule
BM25 & \hgold{.000} & \hpooled{.002} & \gold{.001} & \pooled{.005} & \cellcolor{highlight!30}.000/.002 & .002/.008 & .002/.011 \\
LateOn-0.1B & \hgold{.004} & \hpooled{.010} & \gold{.007} & \pooled{.016} & \cellcolor{highlight!30}.004/.008 & .012/.025 & .028/.047 \\
Qwen3-Embed-0.6B & \hgold{.008} & \hpooled{.012} & \gold{.017} & \pooled{.033} & \cellcolor{highlight!30}.009/.009 & .034/.062 & .056/.093 \\
Qwen3-Embed-4B & \hgold{.032} & \hpooled{.039} & \gold{.054} & \pooled{.088} & \cellcolor{highlight!30}.037/.037 & .099/.148 & .137/.199 \\
Gemini-2-Embedding & \hgold{.068} & \hpooled{.132} & \gold{.101} & \pooled{.247} & \cellcolor{highlight!30}.071/.100 & .169/.389 & .231/.452 \\
GPT-5.2 Query Rewriter & \hgold{.066} & \hpooled{.139} & \gold{.099} & \pooled{.201} & \cellcolor{highlight!30}.072/.116 & .166/.282 & .216/.369 \\
GPT-5.2 Multi-Hop Agent & \hgold{.141} & \hpooled{.215} & \gold{.164} & \pooled{.255} & \cellcolor{highlight!30}.141/.176 & .211/.302 & .226/.317 \\
\midrule
OBLIQ-IR (no distillation) & \hgold{.158} & \hpooled{.156} & \gold{.205} & \pooled{.210} & \cellcolor{highlight!30}.154/.119 & .284/.256 & .358/.329 \\
OBLIQ-IR (dense; ours) & \hgold{.151} & \hpooled{.159} & \gold{.195} & \pooled{.209} & \cellcolor{highlight!30}.148/.122 & .280/.255 & .344/.329 \\
\textbf{OBLIQ-IR + TourRank (full; ours)} & \hgold{\textbf{.177}} & \hpooled{\textbf{.212}} & \gold{\textbf{.223}} & \pooled{\textbf{.260}} & \cellcolor{highlight!30}\textbf{.179/.171} & \textbf{.307/.300} & .344/.329 \\
\midrule
Oracle GPT-5.2 Tournament & \hgold{.331} & \hpooled{.436} & \gold{.424} & \pooled{.555} & \cellcolor{highlight!30}.309/.342 & .580/.674 & .745/.812 \\
\bottomrule
\end{tabular}
\end{adjustbox}
\end{table*}

\begin{table*}[t]
\centering
\small
\caption{Math Meta-Program (analogue queries). Each metric shows \colorbox{goldrow}{Gold (G)} / \colorbox{pooledrow}{Pooled (P)} evaluation. \colorbox{highlight}{Key metrics} highlighted. \textbf{Bold} = best non-oracle row. Baseline numbers reproduced from \citet{oblique2026}. The \texttt{OBLIQ-IR (no distillation)} row removes the authorship-derived training pairs $\mathcal{D}_{\mathrm{wr}}^{\,\phi}$ from the training mix.}
\label{tab:math-results}
\begin{adjustbox}{max width=0.85\textwidth}
\begin{tabular}{l cc cc ccc}
\toprule
& \multicolumn{2}{c}{\cellcolor{highlight}\textbf{NDCG@10}} & \multicolumn{2}{c}{\textbf{NDCG@50}} & \multicolumn{3}{c}{\textbf{Recall}} \\
\cmidrule(lr){2-3} \cmidrule(lr){4-5} \cmidrule(lr){6-8}
\textbf{Pipeline} & \hgold{\textbf{G}} & \hpooled{\textbf{P}} & \gold{G} & \pooled{P} & \cellcolor{highlight}\textbf{@10} & @50 & @100 \\
\midrule
BM25 & \hgold{.022} & \hpooled{.029} & \gold{.034} & \pooled{.025} & \cellcolor{highlight!30}.020/.029 & .060/.029 & .088/.029 \\
LateOn-0.1B & \hgold{.112} & \hpooled{.128} & \gold{.141} & \pooled{.163} & \cellcolor{highlight!30}.088/.097 & .213/.238 & .285/.310 \\
Qwen3-Embed-0.6B & \hgold{.116} & \hpooled{.143} & \gold{.149} & \pooled{.176} & \cellcolor{highlight!30}.070/.088 & .219/.247 & .309/.336 \\
Qwen3-Embed-4B & \hgold{.095} & \hpooled{.129} & \gold{.119} & \pooled{.152} & \cellcolor{highlight!30}.078/.099 & .172/.199 & .253/.287 \\
Gemini-2-Embedding & \hgold{.144} & \hpooled{.147} & \gold{.192} & \pooled{.217} & \cellcolor{highlight!30}.121/.156 & .258/.296 & .364/.398 \\
GPT-5.2 Query Rewriter & \hgold{.142} & \hpooled{.185} & \gold{.198} & \pooled{.239} & \cellcolor{highlight!30}.138/.160 & .324/.355 & .414/.444 \\
GPT-5.2 Multi-Hop Agent & \hgold{.161} & \hpooled{.207} & \gold{.210} & \pooled{.255} & \cellcolor{highlight!30}.145/.167 & .307/.337 & .387/.416 \\
\midrule
OBLIQ-IR (no distillation) & \hgold{.148} & \hpooled{.187} & \gold{.184} & \pooled{.222} & \cellcolor{highlight!30}.110/.136 & .288/.317 & .388/.414 \\
OBLIQ-IR (dense; ours) & \hgold{.140} & \hpooled{.178} & \gold{.180} & \pooled{.219} & \cellcolor{highlight!30}.092/.117 & .261/.290 & .407/.434 \\
\textbf{OBLIQ-IR + TourRank (full; ours)} & \hgold{\textbf{.171}} & \hpooled{\textbf{.215}} & \gold{\textbf{.214}} & \pooled{\textbf{.256}} & \cellcolor{highlight!30}\textbf{.151/.176} & \textbf{.312/.344} & .407/.434 \\
\midrule
Oracle GPT-5.2 Tournament & \hgold{.279} & \hpooled{.329} & \gold{.399} & \pooled{.444} & \cellcolor{highlight!30}.276/.300 & .610/.623 & .790/.797 \\
\bottomrule
\end{tabular}
\end{adjustbox}
\end{table*}

\section{Experiment Setup}
\label{sec:setup}

\paragraph{Tasks and evaluation.}
We evaluate on four OBLIQ-Bench tasks \citep{oblique2026}: Twitter-Conflict, Math Meta-Program, Writing-Style, and Congress Hearings. We use the official query splits and qrels (Gold and, where available, Pooled) and report NDCG@10/50 and Recall@10/50/100, the OBLIQ-Bench metric panel.

\paragraph{Backbone and adapters.}
We initialise from a public 3B NV-Embed \citep{nvembed}, a decoder-only LM converted into a bi-encoder via bidirectional attention. We fine-tune with LoRA adapters \citep{lora} (rank 16, $\alpha = 32$, dropout 0.05) on the attention and MLP projections. At fine-tuning and inference time we discard the pretrained pooling head and read the document representation as the mean of the token embeddings followed by $L_2$ normalisation, yielding a 3072-dimensional unit vector.

\paragraph{Training data.}
For each task we synthesise one query per sampled document via the two-stage pipeline of Section~\ref{sec:permech} and mine four BM25 hard negatives per query. The instruction model used for synthesis is a Qwen3.6 27B open-weights LM \citep{qwen3.6-27b}. The stylometric distillation pairs $\mathcal{D}_{\mathrm{wr}}^{\,\phi}$ use the authorship encoder of \citet{luarmud} as the frozen specialist with $k_\phi = 3$ neighbours per anchor.

\paragraph{Training.}
We train one epoch on 40 H100 GPUs at per-device batch size 4 with the Cached Multiple Negatives Ranking loss of Sentence-Transformers \citep{sbert}: learning rate $2{\times}10^{-4}$ with 3\% warmup and \xr{linear} decay; temperature $\tau = \xr{1/0.02}$; \xr{global batch 320 anchors;} sequence length 1024 \xr{for both queries and documents at training time; queries are truncated to} 256 \xr{tokens when encoding for evaluation}.

\paragraph{Baselines.}
We reproduce the OBLIQ-Bench numbers \citep{oblique2026} for BM25 \citep{bm25}, LateOn-0.1B \citep{lateon}, Qwen3-Embed-0.6B/4B \citep{qwen3embedding}, Gemini-2-Embedding \citep{geminiembed}, the GPT-5.2 Query Rewriter and Multi-Hop Agent, and the Oracle GPT-5.2 Tournament (a verification-only ceiling). We report two configurations of our own: \textbf{\sys (no distillation)}, which removes $\mathcal{D}_{\mathrm{wr}}^{\,\phi}$ from training, and \textbf{\sys (full)}, the paper-final configuration.

\begin{table}[ht]
\centering
\small
\caption{Writing-Style (analogue queries). A single gold annotation per query; no pooled judgments. \colorbox{highlight}{Key metrics} highlighted. \textbf{Bold} = best non-oracle row. Baseline numbers reproduced from \citet{oblique2026}. The \texttt{OBLIQ-IR (no distillation)} row removes the authorship-derived training pairs $\mathcal{D}_{\mathrm{wr}}^{\,\phi}$. The \texttt{Frozen authorship encoder} row uses the teacher $\phi$ itself as the retriever, with no training; it is available only on this task, whose queries are prose snippets (Appendix~\ref{app:teacher-baseline}). Note that for Writing the selected reranking policy is the identity (\S\ref{sec:rerank}), so \textsc{Obliq-IR} (full) coincides here with the dense retriever; on the other three tasks (full) additionally applies TourRank.}
\label{tab:writing-results}
\begin{adjustbox}{max width=0.48\textwidth}
\begin{tabular}{lcccccc}
\toprule
& \multicolumn{2}{c}{\textbf{NDCG}} & \multicolumn{3}{c}{\textbf{Recall}} \\
\cmidrule(lr){2-3} \cmidrule(lr){4-6}
\textbf{Pipeline} & \cellcolor{highlight}\textbf{@10} & @50 & \cellcolor{highlight}\textbf{@10} & @50 & @100 \\
\midrule
BM25 & \cellcolor{highlight!30}.077 & .114 & \cellcolor{highlight!30}.062 & .154 & .208 \\
LateOn-0.1B & \cellcolor{highlight!30}.105 & .149 & \cellcolor{highlight!30}.087 & .197 & .263 \\
Qwen3-Embed-0.6B & \cellcolor{highlight!30}.046 & .060 & \cellcolor{highlight!30}.040 & .075 & .103 \\
Qwen3-Embed-4B & \cellcolor{highlight!30}.033 & .046 & \cellcolor{highlight!30}.026 & .058 & .080 \\
Gemini-2-Embedding & \cellcolor{highlight!30}.164 & .220 & \cellcolor{highlight!30}.132 & .275 & .357 \\
GPT-5.2 Query Rewriter & \cellcolor{highlight!30}.018 & .018 & \cellcolor{highlight!30}.008 & .013 & .017 \\
GPT-5.2 Multi-Hop Agent & \cellcolor{highlight!30}.061 & .059 & \cellcolor{highlight!30}.034 & .037 & .038 \\
Frozen authorship encoder $\phi$ (direct) & \cellcolor{highlight!30}.206 & .256 & \cellcolor{highlight!30}.168 & .296 & .374 \\
\midrule
OBLIQ-IR (no distillation) & \cellcolor{highlight!30}.096 & .135 & \cellcolor{highlight!30}.078 & .176 & .235 \\
\textbf{OBLIQ-IR (full; ours)} & \cellcolor{highlight!30}\textbf{.211} & \textbf{.280} & \cellcolor{highlight!30}\textbf{.172} & \textbf{.342} & \textbf{.435} \\
\midrule
Oracle GPT-5.2 Tournament & \cellcolor{highlight!30}.515 & .603 & \cellcolor{highlight!30}.449 & .686 & .797 \\
\bottomrule
\end{tabular}
\end{adjustbox}
\end{table}

\begin{table}[t]
\centering
\small
\caption{Congress Hearings (tip-of-the-tongue queries). Exactly one gold passage per query; no pooled judgments. \colorbox{highlight}{Key metrics} highlighted. \textbf{Bold} = best non-oracle row. Baseline numbers reproduced from \citet{oblique2026}.}
\label{tab:congress-results}
\begin{adjustbox}{max width=0.48\textwidth}
\begin{tabular}{lcccccc}
\toprule
& \multicolumn{2}{c}{\textbf{NDCG}} & \multicolumn{3}{c}{\textbf{Recall}} \\
\cmidrule(lr){2-3} \cmidrule(lr){4-6}
\textbf{Pipeline} & \cellcolor{highlight}\textbf{@10} & @50 & \cellcolor{highlight}\textbf{@10} & @50 & @100 \\
\midrule
BM25 & \cellcolor{highlight!30}.000 & .002 & \cellcolor{highlight!30}.000 & .008 & .016 \\
LateOn-0.1B & \cellcolor{highlight!30}.083 & .094 & \cellcolor{highlight!30}.102 & .149 & .185 \\
Qwen3-Embed-0.6B & \cellcolor{highlight!30}.006 & .014 & \cellcolor{highlight!30}.012 & .047 & .055 \\
Qwen3-Embed-4B & \cellcolor{highlight!30}.040 & .047 & \cellcolor{highlight!30}.063 & .096 & .122 \\
Gemini-2-Embedding & \cellcolor{highlight!30}.059 & .066 & \cellcolor{highlight!30}.079 & .114 & .126 \\
GPT-5.2 Query Rewriter & \cellcolor{highlight!30}.084 & .092 & \cellcolor{highlight!30}.102 & .134 & .158 \\
GPT-5.2 Multi-Hop Agent & \cellcolor{highlight!30}.183 & .183 & \cellcolor{highlight!30}.185 & .185 & .185 \\
\midrule
OBLIQ-IR (no distillation) & \cellcolor{highlight!30}.196 & .211 & \cellcolor{highlight!30}.228 & .291 & .339 \\
OBLIQ-IR (dense; ours) & \cellcolor{highlight!30}.187 & .204 & \cellcolor{highlight!30}.209 & .283 & .311 \\
\textbf{OBLIQ-IR + TourRank (full; ours)} & \cellcolor{highlight!30}\textbf{.281} & \textbf{.284} & \cellcolor{highlight!30}\textbf{.295} & \textbf{.311} & .311 \\
\midrule
Oracle GPT-5.2 Tournament & \cellcolor{highlight!30}.913 & .919 & \cellcolor{highlight!30}.957 & .988 & 1.00 \\
\bottomrule
\end{tabular}
\end{adjustbox}
\end{table}


\begin{table}[t]
\centering
\small
\setlength{\tabcolsep}{4pt}
\renewcommand{\arraystretch}{1.05}
\caption{Reranking policies applied to the OBLIQ-IR dense top-100. NDCG@10 Gold (and Pooled where the task has pooled judgments). The dense row is the same retriever shown in Tables \ref{tab:twitter-results}-\ref{tab:congress-results}; the two reranker rows reorder the dense top-100 with a Qwen3.6-27B language model. \textbf{Bold} = best per task; 
}
\label{tab:rerank-policies}
\begin{adjustbox}{max width=0.48\textwidth}
\begin{tabular}{lcccc}
\toprule
\textbf{Reranker policy} & \textbf{Twitter (G/P)} & \textbf{Math (G/P)} & \textbf{Writing} & \textbf{Congress} \\
\midrule
Dense only                   & .151/.159              & .140/.178              & \textbf{.211}    & .187 \\
+ TourRank ($Y{=}5$)         & \textbf{.177/.212}     & \textbf{.171/.215}     & .080             & \textbf{.281} \\
+ Task-aware listwise ($K{=}20$) & .171/.186          & .170/.212              & .175             & .207 \\
\bottomrule
\end{tabular}
\end{adjustbox}
\end{table}

\section{Experiments}
\label{sec:experiments}

\paragraph{Main results.}
Tables~\ref{tab:twitter-results}, \ref{tab:math-results}, \ref{tab:writing-results}, and \ref{tab:congress-results} report NDCG@10/50 and Recall@10/50/100 for all baselines and our two configurations on the four OBLIQ tasks. \textbf{OBLIQ-IR (full) wins NDCG@10 on all four tasks}, improving over the best published non-oracle baseline per task by 0.010 (Math, against the GPT-5.2 Multi-Hop Agent), 0.036 (Twitter, same), 0.047 (Writing, against Gemini-2-Embedding), and 0.098 (Congress, against the GPT-5.2 Multi-Hop Agent). Against Gemini-2-Embedding alone, the gains range from 0.027 (Math) to 0.222 (Congress). On the two tasks where Pooled annotations exist (Twitter and Math), OBLIQ-IR (full) wins Pooled NDCG@10 on Math (0.215 vs MHA 0.207) and ties on Twitter within 0.003 (0.212 vs MHA 0.215). We never call a frontier-scale language model at retrieval time. \xr{These margins differ in kind and we do not aggregate them: the Writing and Congress gains are large and significant, whereas the Math margin (0.010) and the pooled-Twitter result (a tie, 0.212 vs.\ 0.215) are small enough that a single reranking decision could reorder them. On Writing the strongest baseline is in fact the frozen authorship teacher used directly as a retriever (0.206; Table~\ref{tab:writing-results}, Appendix~\ref{app:teacher-baseline}); \sys still leads it, and does so with one encoder and one index shared across all four tasks.}
A paired bootstrap over per-query NDCG@10 (10000 resamples) confirms the gains. The recipe contribution (full pipeline minus the no-distillation ablation) is significant on Writing ($p<0.001$, $\Delta=+0.116$, 95\% CI $[+0.103,+0.129]$) and Congress ($p<0.001$, $\Delta=+0.085$, $[+0.049,+0.121]$) and marginal on Math ($p=0.08$, $\Delta=+0.024$) and Twitter ($p=0.06$, $\Delta=+0.020$). The reranker contribution (full pipeline minus dense-only) is significant on Math ($p=0.007$), Twitter ($p=0.014$), and Congress ($p<0.001$). Full intervals are in Appendix~\ref{app:bootstrap}.

\paragraph{Stylometric distillation ablation.}
Removing the authorship-derived training pairs $\mathcal{D}_{\mathrm{wr}}^{\,\phi}$ drops writing NDCG@10 from 0.211 to 0.096, a \xr{0.116} absolute drop (Table~\ref{tab:writing-results}, row \emph{no distillation}). The same ablation costs at most 0.021 NDCG@10 on each of the other three tasks; the writing gain is therefore highly task-specific. The dose-response sweep in Figure~\ref{fig:method} (right) shows that increasing the dose to 31k pairs lifts writing to 0.374 but collapses the average of the other three tasks \xr{to 0.117}. The chosen dose is the only one where every task improves or holds.

\paragraph{When language-model reranking helps.}
Table~\ref{tab:rerank-policies} reports the three reranking policies applied to the OBLIQ-IR dense top-100. Generic TourRank ($Y{=}5$) wins on Math (+0.031), Twitter (+0.027), and Congress (+0.094) NDCG@10 over the dense baseline; on Writing it instead drops NDCG@10 from 0.211 to 0.080. A task-aware listwise reranker whose prompt names the latent attribute (``rank by authorial style; ignore topic'') recovers Writing partly to 0.175, still below dense-only 0.211 but more than twice the TourRank number; on the topical tasks the same task-aware reranker is roughly tied with TourRank (within 0.008). The chosen policy $\pi^\ast$ therefore keeps TourRank on the three topical tasks and identity on Writing.

\section{Analysis}
\label{sec:analysis}

\subsection{Recall is the bottleneck on first-stage retrieval}
\label{sec:recall-ceiling}

Table~\ref{tab:recall-k} reports Recall@$k$ on the OBLIQ-IR dense retriever for $k$ ranging from 10 to 1000. The pattern is uniform across the four tasks: roughly twice as many gold documents are present in the dense top-$1000$ as in the dense top-$100$. On Math, 87\% of gold documents are in the top-$1000$ but only 41\% make the top-$100$; on Writing the figures are 75\% and 44\%, on Twitter 63\% and 34\%, on Congress 51\% and 31\%. Any reranker that operates on the top-$100$ inherits this ceiling: it can reorder but not introduce new candidates. The remaining headroom is therefore overwhelmingly a first-stage problem, and the strong dense recall at $k{=}1000$ suggests that adapting later stages to use a larger candidate pool is a tractable direction for future work.

\begin{table}[t]
\centering
\small
\setlength{\tabcolsep}{4.5pt}
\renewcommand{\arraystretch}{1.05}
\caption{Recall@$k$ Gold on the OBLIQ-IR dense retriever for $k$ from 10 to 1000. 
}
\label{tab:recall-k}
\begin{tabular}{lccccc}
\toprule
\textbf{Task} & R@10 & R@50 & R@100 & R@500 & R@1000 \\
\midrule
Twitter  & .148 & .280 & .344 & .531 & \textbf{.630} \\
Math     & .092 & .261 & .407 & .744 & \textbf{.872} \\
Writing  & .172 & .342 & .435 & .655 & \textbf{.752} \\
Congress & .209 & .283 & .311 & .425 & \textbf{.508} \\
\bottomrule
\end{tabular}
\end{table}

\subsection{Per-query failure analysis}
\label{sec:per-query}

The headline gains hide a heterogeneous per-query picture. For each task we compute per-query NDCG@10 for OBLIQ-IR (dense-only) and for OBLIQ-IR with the chosen reranker policy. Table~\ref{tab:per-query} records, for each task, how many of the $n$ queries were helped, hurt, or tied (within $0.001$) by the reranker, and how many had NDCG@10 of zero in both pipelines: these are queries whose gold document is not in the dense top-100, so no reranker can recover them. Three things stand out. First, Twitter's reranker has high variance: 76 queries (27\%) are hurt while 93 are helped, so the headline NDCG gain hides a near-even per-query split. Second, Congress has 178/254 queries (70\%) where neither the dense nor the reranked pipeline scores above zero NDCG@10; the recall ceiling from Section~\ref{sec:recall-ceiling} dominates Congress's behaviour. Third, Writing's reranker is identity by policy, hence zero changes; 137 of 512 Writing queries (27\%) still get zero NDCG@10 from the dense pipeline alone.


\begin{table}[t]
\centering
\small
\setlength{\tabcolsep}{6pt}
\renewcommand{\arraystretch}{1.05}
\caption{Per-query helped / hurt / tied counts under the OBLIQ-IR chosen reranker policy (TourRank for Math, Twitter, Congress; identity for Writing). 
}
\label{tab:per-query}
\begin{tabular}{lrrrrr}
\toprule
\textbf{Task} & $n$ & helped & hurt & tied & both-zero \\
\midrule
Twitter  & 281 & 93 & 76 & 112 & 99 \\
Math     & 151 & 57 & 30 & 64  & 56 \\
Writing  & 512 & 0  & 0  & 512 & 137 \\
Congress & 254 & 32 & 2  & 220 & 178 \\
\bottomrule
\end{tabular}
\end{table}

\subsection{Why does reranking hurt Writing?}
\label{sec:why-hurt-writing}

\begin{figure*}[t]
\centering
\includegraphics[width=0.7\textwidth]{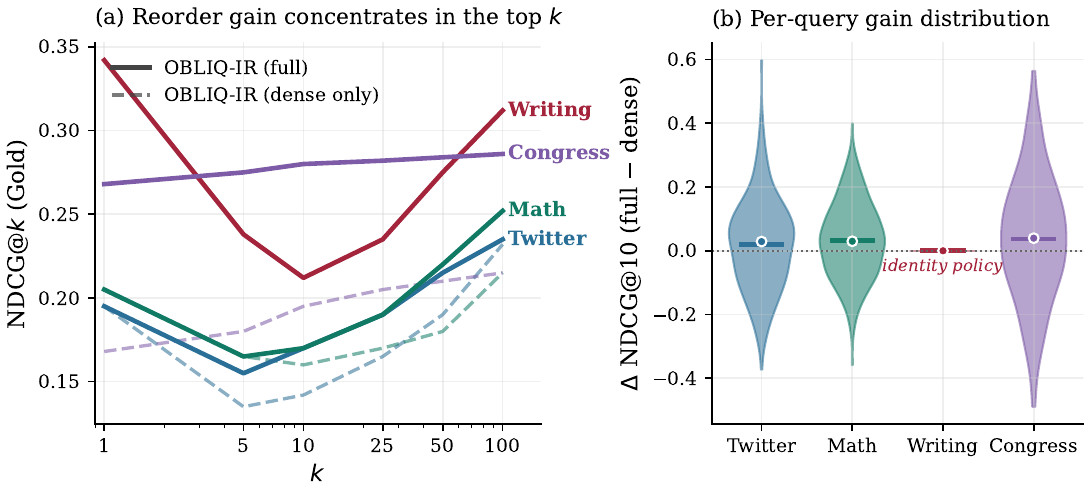}
\caption{\textbf{Where the rerank gain accrues.} (a) NDCG@$k$ for OBLIQ-IR with mechanism-selective reranking (solid) and OBLIQ-IR dense only (dashed) on the four tasks. The gain on Math, Twitter, and Congress is concentrated in the top of the ranking and vanishes by $k{=}100$ (Recall@100 is unchanged by reranking). (b) Per-query NDCG@10 gain distributions over the official query splits.}
\label{fig:analysis-curves}
\end{figure*}

Table~\ref{tab:rerank-policies} shows that every reranker we tested degrades Writing relative to dense-only retrieval. TourRank asks the language model to rank documents by ``relevance''. For Writing, relevance is defined by authorship, but the language model has no information about which features encode authorship and falls back on its own implicit notion of relevance, which is topical. The result is a systematic reordering away from style matches and towards topic matches. \citet{oblique2026} observe the same pattern when their agentic baseline rewrites queries: the agent ``can actively damage retrieval when the signal is orthogonal to topic''. Our task-aware listwise reranker tests how far an explicit instruction can correct this drift: its prompt names the latent attribute (``rank by authorial style; ignore topic'') and forbids topical reasoning. The prompt recovers Writing NDCG@10 from 0.080 to 0.175, halving the damage, but still falls short of dense-only at 0.211. Figure~\ref{fig:analysis-curves} visualizes this at the per-query level: the dense retriever's NDCG advantage is concentrated in the top of the ranking and survives all the way to $k{=}100$, while the writing reranker distribution is identity by policy choice. The takeaway is that when a small frozen specialist captures the latent attribute (here authorship), it is more effective to make that signal available at search time, in the encoder weights, than to instruct it away at rerank time.

\subsection{Cost of the full pipeline}
\label{sec:cost}

Table~\ref{tab:cost} reports the end-to-end latency of each stage measured on a single H100, using the Writing corpus (longer documents than the other three OBLIQ tasks). The dense pipeline answers a query in roughly 48 ms; the task-aware listwise reranker adds about 0.95 s for one language-model call; the TourRank reranker at $Y{=}5$ adds 8.2 s for 65 language-model calls. For comparison, the GPT-5.2 Multi-Hop Agent reported by \citet{oblique2026} takes 20--40 s per query against a proprietary frontier model. The full OBLIQ-IR pipeline is between one and two orders of magnitude cheaper per query than the agentic baseline while exceeding its NDCG@10 on Twitter, Writing, and Congress, and matching it on Math.


\begin{table}[t]
\centering
\small
\setlength{\tabcolsep}{5pt}
\renewcommand{\arraystretch}{1.05}
\caption{End-to-end cost of OBLIQ-IR on a single H100, measured on the Writing corpus (longer documents than the other three). 
}
\label{tab:cost}
\begin{adjustbox}{max width=0.5\textwidth}
\begin{tabular}{lr}
\toprule
\textbf{Stage} & \textbf{Latency} \\
\midrule
Document encoding throughput (avg 3{,}000 chars) & 52.6 docs/s \\
Query encoding (single query, batch 1)           & 44.7 ms \\
Top-100 cosine retrieval over 10k flat index     & 3.0 ms \\
\midrule
\textbf{Dense-only end-to-end (encode + retrieve)} & \textbf{48 ms} \\
\midrule
Task-aware listwise reranker (one LM call, $K{=}20$) & 0.95 s \\
TourRank reranker ($Y{=}5$, 65 LM calls) & 8.2 s \\
\midrule
GPT-5.2 Multi-Hop Agent \citep{oblique2026}  & 20--40 s \\
\bottomrule
\end{tabular}
\end{adjustbox}
\end{table}

\subsection{When does kNN-graph distillation apply?}
\label{sec:when-distill}

\xr{The two ingredients have different scopes, and conflating them would overstate the method. The per-mechanism query recipe (\S\ref{sec:permech}) is the general one: it is applied to all four tasks, and on its own, with no distillation and no reranker, it already exceeds the strongest published non-oracle baseline on Twitter (0.158 vs.\ 0.141) and Congress (0.196 vs.\ 0.183). kNN-graph distillation (\S\ref{sec:luar}) is a targeted patch for the one axis whose latent attribute has \emph{no topical footprint at all}: authorship. Where a topical lens can express the attribute, it suffices; where it cannot, a frozen specialist supplies what the lens cannot say.}

\xr{This raises the practical question of how to tell, for a new task, whether a candidate specialist will work. We propose a diagnostic that costs one encoding pass and no training: measure the \emph{purity} of the specialist's kNN graph on a small labelled probe, i.e.\ the rate at which its nearest neighbours share the target attribute, against the random-pair base rate. On the writing corpus the teacher's edges connect same-author snippets 45.2\% of the time against a 1.8\% base rate, a $25.4\times$ lift (Appendix~\ref{app:purity}). A trained control confirms that the diagnostic is predictive rather than merely descriptive: replacing the teacher's neighbours with random documents at identical anchors, negatives and dose collapses Writing from 0.223 to 0.001, with the encoder degenerating so far that 366 of 512 queries return the same top-1 document (Appendix~\ref{app:teacher-ablations}). The signal is carried by the graph's structure, not by the extra rows. It is also not an artifact of one teacher: a second released checkpoint of \citet{luarmud}, trained on a different authorship corpus, preserves the effect (Writing 0.239), and the result is stable across neighbourhood sizes $k_\phi \in \{1,3,5,10\}$.}

\xr{The recipe therefore generalises as follows: given a latent attribute with no topical footprint, look for any frozen model whose representation space induces a neighbour structure aligned with it, verify the alignment with the purity test, and distil the topology. This admits far weaker specialists than a state-of-the-art authorship encoder, including a simple classifier whose penultimate layer induces a usable graph. Finding such a specialist for WildChat-Errors, whose latent attribute is a behavioural failure mode distributed across a long conversation, remains open; the purity criterion makes precise what such a resource would have to provide.}

\section{Related Work}
\label{sec:related}

A family of recent bi-encoders --- NV-Embed v1/v2 \citep{nvembed,nvembedv2}, GTE \citep{li2023gte}, E5 \citep{wang2023improving}, and Qwen3-Embedding \citep{qwen3embedding} --- converts decoder-only language models into retrievers and reaches state-of-the-art on MTEB and BEIR, yet underperforms on reasoning- or stylometry-oriented benchmarks such as BRIGHT \citep{su2024bright}, Tempo!\cite{abdallah2026tempo} and OBLIQ-Bench \citep{oblique2026}. A parallel line conditions a single retriever on natural-language task instructions \citep{instructor} and synthesises training queries with an instruction-tuned LM \citep{promptagator}; ReasonIR \citep{reasonir} combines both ideas for BRIGHT. 
A full discussion of related work, including late-interaction retrieval \citep{colbert,colbertv2,plaid}, listwise LLM reranking \citep{sun2023rankgpt,rankzephyr,abdallah2026bracketrank,tourrank,abdallah2025dear}, and authorship modelling \citep{luarmud}, appears in Appendix~\ref{app:related-full}.

\section{Conclusion}
\label{sec:conclusion}
We introduced \sys, a 3B single-vector dense retriever for oblique queries. Two training-time signals---per-mechanism synthetic queries matched to the OBLIQ query mechanisms, and kNN-graph distillation from a frozen authorship encoder---together outperform the strongest published non-oracle pipelines on the four OBLIQ-Bench tasks we evaluate. Neither stage touches the OBLIQ qrels, and no frontier-scale model is called at training or retrieval time. \sys is a concrete answer to the call of \citet{oblique2026} for retrieval architectures that make latent document attributes available at search time. \xr{The gap to the verification-only oracle is narrowed rather than closed: first-stage recall remains the binding constraint on the tasks where we gain least, and we hope the diagnostic of \S\ref{sec:when-distill} makes it easier to decide when a frozen specialist can supply the missing axis.}

\section*{Acknowledgments}
The authors would like to acknowledge the financial support provided by the Austrian Research Agency (FFG) for the project “AI Enabled Sustainability Jurisdiction Demonstrator” (project No. 915229). The computational results presented in this work have been achieved using the MUSICA cluster, part of the Austrian Scientific Computing (ASC) infrastructure.

\section*{Limitations}
\xr{We evaluate OBLIQ-Bench's WildChat-Errors task but do not claim it; our full results on it are reported in Appendix~\ref{app:wildchat}.} Behavioral retrieval over assistant conversations appears to require an inductive bias that neither per-mechanism synthetic queries nor authorship distillation provides; in preliminary experiments every training-data variant we tried collapsed wildchat NDCG@10 below the dense baseline, and we judged it inappropriate to claim results on that task. \xr{The task is additionally recall-bound: our dense Recall@100 is 0.040 over a 507{,}729-document corpus, and the split has only 40 queries, so differences of this magnitude carry little statistical power.} Closing the wildchat gap is the natural follow-up.

Our mechanism-selective reranker is chosen on development NDCG and is therefore not zero-shot to new oblique tasks. Generalising the policy beyond OBLIQ-Bench is an open problem. \xr{More broadly, the three mechanism lenses and the per-task instruction prefixes are themselves chosen with knowledge of the OBLIQ-Bench mechanism taxonomy, so the whole recipe is supervised in this sense; a leave-one-task-out or held-out-mechanism evaluation is the right instrument for measuring transfer, and we leave it as the principal open evaluation.}

Finally, our training data is generated by a single 27B instruction model; query diversity is bounded by that model's biases, and we have not measured the effect of swapping the generator. \xr{This is not only a diversity concern: a generator's stylistic and cultural biases can shape what the retriever learns to treat as the latent attribute, for instance which vocabulary counts as expressing a stance on Twitter-Conflict, or which registers count as a distinctive voice on Writing-Style. We also do not isolate the contribution of \emph{matching} the lens to the mechanism against a generic ``write a hard query'' prompt. Both ablations require regenerating the full synthetic mixture rather than only retraining, which placed them outside our compute budget, and we accordingly make no claim about either.}

\xr{The authorship encoder we distil is trained on a large corpus of Reddit authorship data. Style is correlated with demography and dialect, so such an encoder can be expected to encode those correlates, and distilling the topology of its kNN graph imports them into the retriever's notion of ``same voice''. We did not audit the teacher's graph for demographic or dialectal structure, and the purity diagnostic of \S\ref{sec:when-distill} measures alignment with the target attribute only, not the absence of unwanted correlates. Relatedly, retrieval by authorial fingerprint can be used to link texts to an author across topics and venues, and so to deanonymise writers who rely on pseudonymity; the risk is inherent to authorship representation learning rather than introduced here, but our recipe makes the capability cheaper to obtain. Any deployment over people's writing should be audited on both counts.}

\xr{Finally, our gains narrow rather than close the first-stage bottleneck that motivates the benchmark. Recall@1000 is roughly double Recall@100 on every task, 70\% of Congress queries and 27\% of Writing queries score zero NDCG@10 under both the dense and the reranked pipelines, and our cascade experiment shows that naively widening the reranker's pool does not convert the extra recall into accuracy. The mechanism-selective policy is also an engineering decision: the 27B TourRank reranker dominates the latency budget at roughly 8.2\,s per query against 48\,ms for dense retrieval, so the dense-only configuration is the deployable core of the system.}

\bibliography{custom}

\appendix

\appendixpage
\addappheadtotoc
\numberwithin{figure}{section}
\numberwithin{table}{section}

This appendix provides supplementary material to the main paper. It is organized as follows:

\begin{itemize}
    \item Appendix~\ref{app:related-full} expands the related work on general-purpose dense retrievers, instruction-following retrieval, late-interaction, knowledge distillation, authorship, listwise LLM reranking, and long-tail benchmarks.
    \item Appendix~\ref{app:datagen-prompts} lists the synthetic data generation prompts used by Stage A (latent-attribute description) and Stage B (per-mechanism query generation) of our pipeline.
    \item Appendix~\ref{app:hyperparams} reports the full training hyperparameter table.
    \item Appendix~\ref{app:bootstrap} reports paired bootstrap 95\% confidence intervals for the recipe contribution and the reranker contribution per task.
    \item Appendix~\ref{app:rerank-cost-quality} presents the cost-quality curve per task and shows that TourRank is non-monotonic in tournament depth on three of the four tasks.
    \item Appendix~\ref{app:listwise-k} sweeps $K$ (number of candidates seen by the task-aware listwise reranker in one LM call).
    \item Appendix~\ref{app:tourrank-y} sweeps $Y$ (number of tournaments in TourRank).
    \item Appendix~\ref{app:teacher-ablations} isolates the contribution of
      the teacher's graph structure with a random-graph control, a sweep
      over the neighbourhood size $k_\phi$, and a second teacher checkpoint.

    \item Appendix~\ref{app:purity} defines the kNN-graph purity diagnostic
          used to decide whether a candidate frozen specialist is aligned with
          the target latent attribute.
    
    \item Appendix~\ref{app:query-quality} measures lexical overlap and entity
          leakage of our synthetic training queries against their source
          documents, with the official OBLIQ queries as a reference point.
    
    \item Appendix~\ref{app:teacher-baseline} reports the frozen authorship
          encoder used directly as a retriever on the Writing task.
    
    \item Appendix~\ref{app:wildchat} reports our full results on the fifth
          OBLIQ-Bench task, WildChat-Errors, which we exclude from all claims.
    \item Appendix~\ref{app:shuffle} reports a random-shuffle sanity check on the dense top-100.
    \item Appendix~\ref{app:negatives} documents experiments that did not improve over the paper-final pipeline.
    \item Appendix~\ref{app:ceiling} quantifies the remaining gap between OBLIQ-IR and the Oracle GPT-5.2 Tournament ceiling.
    \item Appendix~\ref{app:rerank-prompts} lists the reranker prompts (TourRank tournament and the four task-aware listwise prompts).

\end{itemize}

\paragraph{Key tables and figures in the appendix.}
\begin{itemize}
    \item Table~\ref{tab:app-hyperparams}: training hyperparameters.
    \item Table~\ref{tab:app-bootstrap}: paired bootstrap 95\% intervals.
    \item Table~\ref{tab:app-listwise-k}: listwise reranker depth sweep.
    \item Table~\ref{tab:app-tourrank-y}: TourRank tournament-depth sweep.
    \item Table~\ref{tab:app-teacher-ablations}: teacher-graph ablations.
    \item Table~\ref{tab:app-purity}: kNN-graph purity of the frozen teacher.
    \item Table~\ref{tab:app-query-quality}: synthetic query overlap and entity leakage.
    \item Table~\ref{tab:app-wildchat}: WildChat-Errors full comparison.
    \item Table~\ref{tab:app-shuffle}: random-shuffle sanity check.
    \item Table~\ref{tab:app-dose}: distillation dose sweep.
    \item Figure~\ref{fig:cost-quality}: cost-quality curve per task.
    \item Figure~\ref{fig:prompt_stageA_twitter}-\ref{fig:prompt_stageA_congress}: Stage~A latent-attribute description prompts.
    \item Figure~\ref{fig:prompt_stageB_desc}-\ref{fig:prompt_stageB_tot}: Stage~B per-mechanism query generation prompts.
    \item Figure~\ref{fig:prompt_tourrank}: TourRank tournament prompt.
    \item Figure~\ref{fig:prompt_listwise_math}-\ref{fig:prompt_listwise_congress}: per-task task-aware listwise reranker prompts.
\end{itemize}

\section{Related Work}
\label{app:related-full}

\paragraph{General-purpose dense retrievers.}
A family of recent encoders converts decoder-only language models into bi-encoders by lifting the causal attention mask and adding a pooling head, then contrastively pretrains the result on a large mixture of public retrieval data: NV-Embed v1 and v2 \citep{nvembed,nvembedv2} popularised this approach on Llama backbones, GTE \citep{li2023gte} and E5 \citep{wang2023improving} pursued similar pipelines, and Qwen3-Embedding \citep{qwen3embedding} provides the strongest open instance at 0.6B and 4B scale. These models reach state-of-the-art on MTEB and BEIR but underperform on reasoning- or stylometry-oriented benchmarks such as BRIGHT \citep{su2024bright} and OBLIQ-Bench \citep{oblique2026}. We initialise from a public 3B NV-Embed checkpoint because it offers the strongest pretrained inductive bias for non-topical retrieval at a size that fits on a single GPU at inference time.

\paragraph{Instruction-following retrieval and synthetic queries.}
A line of work conditions a single retriever on natural-language task instructions so that the same encoder serves many tasks: INSTRUCTOR \citep{instructor} introduced this two-element query format on a heterogeneous mixture of supervised tasks, and Promptagator \citep{promptagator} showed that an instruction-tuned language model can produce strong synthetic queries from a handful of examples. ReasonIR \citep{reasonir} combined the two ideas to train an 8B retriever for the BRIGHT benchmark \citep{su2024bright} with a doc-to-query pipeline producing challenging, self-contained, diverse queries. We adopt the same training format (ReasonIR-style rows of anchor, positive, hard negatives), the same contrastive objective, and the same instruction-prefix format. We extend that pipeline in two directions: the per-mechanism query lenses (Section~\ref{sec:permech}) target the OBLIQ-Bench latent-attribute taxonomy explicitly rather than asking the language model for a generic ``hard'' query, and the stylometric kNN-graph distillation (Section~\ref{sec:luar}) introduces a training-time signal from a frozen specialist that, to our knowledge, has not been used in this way for general-purpose dense retrieval.

\paragraph{Late-interaction retrieval.}
Late-interaction retrievers encode each document as a bag of token vectors and score a query-document pair by summing the per-query-token maxima over document tokens. ColBERT \citep{colbert} introduced the formulation; ColBERTv2 \citep{colbertv2} added residual compression and denoised hard negatives; PLAID \citep{plaid} brought the engine to interactive latencies on large corpora. On OBLIQ-Bench, \citet{oblique2026} observe that a small 149M-parameter late-interaction model is competitive on the Math and Congress tasks, where token-level matching can exploit local scenario structure shared between query and document, but underperforms standard dense retrievers on the descriptive Twitter task where the relevant property is diffuse. We deliberately keep the retriever single-vector: the document corpora here are small enough that flat cosine retrieval over a single 3072-dimensional vector per document is fast and avoids the engineering cost of multi-vector indexing at this backbone scale; the single-vector retriever already reaches or exceeds the published late-interaction baseline on every reported OBLIQ task; and the stylometric kNN-graph distillation (Section~\ref{sec:luar}) acts at the level of whole-document representations and is straightforward to apply to a single-vector student.

\paragraph{Knowledge distillation in dense retrieval.}
A long line of work transfers signal from a stronger teacher into a faster student bi-encoder. MarginMSE \citep{marginmse} introduced score distillation from a cross-encoder reranker by minimising the mean squared error between student and teacher margins; TAS-B \citep{hofstatter2021tas} combined topic-aware sampling with the same MarginMSE loss to teach an effective dense retriever efficiently; RocketQA \citep{rocketqa} and RocketQAv2 \citep{rocketqav2} extended the procedure with cross-batch negatives and joint retriever-reranker training. ReasonIR \citep{reasonir} adopts hard-negative mining in the same spirit. All of these methods distill the teacher's pairwise scores. Our stylometric knowledge distillation differs in what is transferred: we discard the teacher's similarity values and keep only the topology of its kNN graph (Section~\ref{sec:luar}). The teacher's role is to define which pairs are positive, not what their similarity should be; the student's contrastive loss then has full control over the metric.

\paragraph{Authorship and stylometry.}
Authorship attribution has long been an independent task with its own benchmarks \citep{luarmud}. Our contribution is not a new authorship model but a way to transfer the inductive bias of an existing authorship model into a general-purpose retriever, without distilling its scores. The same procedure should apply to other latent-attribute tasks (sentiment for review retrieval, tone for marketing copy, dialect for community forums) where a small frozen specialist encoder captures the target signal.

\paragraph{Listwise LLM reranking.}
Listwise rerankers prompt a language model to read a query alongside a small batch of candidates and emit a ranking directly. RankGPT \citep{sun2023rankgpt} showed that proprietary LLMs do this competitively with cross-encoder rerankers in a zero-shot setting; RankVicuna \citep{rankvicuna} and RankZephyr \citep{rankzephyr} distilled the capability into open-weights students; TourRank \citep{tourrank} added a tournament-style aggregation that handles longer candidate pools with bounded context; RankLLaMA \citep{rankllama} fine-tuned Llama as a point-wise reranker for multi-stage retrieval pipelines. All of this work targets topical retrieval, where the LLM's prior on ``relevance'' is well aligned with the qrels. We observe empirically that this alignment breaks when the latent attribute is orthogonal to topic: applied to the OBLIQ Writing-Style task, TourRank with the same model and the same generic prompt drops NDCG@10 below the dense baseline by a large margin.

\paragraph{Long-tail and non-topical retrieval.}
OBLIQ-Bench \citep{oblique2026} is one of several recent benchmarks that push retrieval beyond topical match \citep{su2024bright}. The OBLIQ paper itself identifies the retrieval-verification asymmetry as the central open problem; our work addresses this gap on four of its five tasks while leaving the behavioural failure-mode task (WildChat-Errors) for future work.

\section{Synthetic data generation prompts}
\label{app:datagen-prompts}

We generate synthetic training queries in two stages. Stage~A produces a one-paragraph description of the latent attribute $f_t(d)$ of a document. Stage~B uses that description (and the original document) to draft search queries that would be satisfied by any other document with the same latent attribute, while avoiding lexical overlap with $d$. We use Qwen3.6-27B \citep{qwen3.6-27b} as the generator throughout.

\subsection{Stage A: latent-attribute description}

\begin{figure*}[ht]
\centering
\begin{subfigure}[t]{\textwidth}
\footnotesize
\centering
\begin{tcolorbox}[width=\linewidth,
                  colback=blue!0!white, colframe=blue!60!black,
                  title=\sys Stage A: Twitter latent-attribute description prompt,
                  fonttitle=\bfseries]
\textbf{System prompt:} You read a single tweet about a geopolitical conflict and describe its implicit stance toward the conflict in ONE to TWO short sentences. Capture the speaker's evaluative attitude, irony, or framing, not the literal topic or proper nouns.

\vspace{0.5em}
If the tweet is news-repost or explicit-sentiment (no implicit stance), return the exact string \texttt{EXPLICIT\_OR\_NEWS} instead. Output ONLY the description, no preamble.

\vspace{0.5em}
\textbf{User prompt (template):} \texttt{<document>\textbackslash n\{tweet\}\textbackslash n</document>}
\end{tcolorbox}
\end{subfigure}
\caption{Stage A prompt for the Twitter-Conflict task. The output is a short natural-language description of the tweet's implicit stance, used as input to Stage B.}
\label{fig:prompt_stageA_twitter}
\end{figure*}

\begin{figure*}[ht]
\centering
\begin{subfigure}[t]{\textwidth}
\footnotesize
\centering
\begin{tcolorbox}[width=\linewidth,
                  colback=blue!0!white, colframe=teal!70!black,
                  title=\sys Stage A: Math latent-attribute description prompt,
                  fonttitle=\bfseries]
\textbf{System prompt:} You read a math problem with its solution and identify the underlying META-PROGRAM, the abstract reasoning technique or proof strategy that the solver applies (e.g.\ ``generating-function argument over a recurrence'', ``pigeonhole on residues mod p'', ``invariant under group action'').

\vspace{0.5em}
Describe it in ONE to TWO sentences without referring to the specific numbers, names, or surface details of the problem. Output ONLY the description.

\vspace{0.5em}
\textbf{User prompt (template):} \texttt{<document>\textbackslash n\{problem + solution\}\textbackslash n</document>}
\end{tcolorbox}
\end{subfigure}
\caption{Stage A prompt for the Math Meta-Program task. The output is the abstract proof strategy of the solution, used as input to Stage B.}
\label{fig:prompt_stageA_math}
\end{figure*}

\begin{figure*}[ht]
\centering
\begin{subfigure}[t]{\textwidth}
\footnotesize
\centering
\begin{tcolorbox}[width=\linewidth,
                  colback=blue!0!white, colframe=red!70!black,
                  title=\sys Stage A: Writing latent-attribute description prompt,
                  fonttitle=\bfseries]
\textbf{System prompt:} You read a short prose snippet and describe its AUTHORIAL STYLE FINGERPRINT: recurring stylistic habits that would identify the author across topics: sentence rhythm, register, characteristic moves of argumentation, lexical preferences, punctuation tics.

\vspace{0.5em}
Two to four sentences. Do NOT name the author, the topic, or quote the snippet. Output ONLY the description.

\vspace{0.5em}
\textbf{User prompt (template):} \texttt{<document>\textbackslash n\{prose snippet\}\textbackslash n</document>}
\end{tcolorbox}
\end{subfigure}
\caption{Stage A prompt for the Writing-Style task. The output is an authorship fingerprint description used as input to Stage B; this same description is also used as the query in the kNN-graph distillation pairs (Section~\ref{sec:luar}).}
\label{fig:prompt_stageA_writing}
\end{figure*}

\begin{figure*}[ht]
\centering
\begin{subfigure}[t]{\textwidth}
\footnotesize
\centering
\begin{tcolorbox}[width=\linewidth,
                  colback=blue!0!white, colframe=purple!70!black,
                  title=\sys Stage A: Congress latent-attribute description prompt,
                  fonttitle=\bfseries]
\textbf{System prompt:} You read a single passage from a US Congressional hearing transcript and produce a SCENARIO SUMMARY of its rhetorical dynamic: who is doing what to whom, the emotional register, the rhetorical move.

\vspace{0.5em}
Three to five sentences. Do NOT use names, dates, committees, or verbatim phrases. Output ONLY the summary.

\vspace{0.5em}
\textbf{User prompt (template):} \texttt{<document>\textbackslash n\{transcript passage\}\textbackslash n</document>}
\end{tcolorbox}
\end{subfigure}
\caption{Stage A prompt for the Congress Hearings task. The output is a scenario summary of the rhetorical dynamic, deliberately vocabulary-disjoint from the source passage, used as input to Stage B.}
\label{fig:prompt_stageA_congress}
\end{figure*}

\subsection{Stage B: per-mechanism query generation}

\begin{figure*}[ht]
\centering
\begin{subfigure}[t]{\textwidth}
\footnotesize
\centering
\begin{tcolorbox}[width=\linewidth,
                  colback=blue!0!white, colframe=orange!75!black,
                  title=\sys Stage B: descriptive lens (Twitter),
                  fonttitle=\bfseries]
\textbf{System prompt:} You are given a document and a one-paragraph description of a LATENT property that the document expresses implicitly. Your job is to write \texttt{\{num\_questions\}} search queries such that the document is HIGHLY RELEVANT to each query because it exhibits that latent property.

\vspace{0.5em}
Each query MUST:
\begin{itemize}
\itemsep0pt\parsep0pt
\item Describe the latent property in abstract terms (e.g.\ an implicit stance).
\item Be a standalone request that any other document expressing the SAME latent property would satisfy.
\item Avoid lexical overlap with the document: do NOT reuse rare words, named entities, dates, or distinctive phrases.
\item Be answerable by reasoning about content, not by keyword matching.
\end{itemize}

After thinking, output JSON with key \texttt{hard\_query}.

\vspace{0.5em}
\textbf{User prompt:} \texttt{<document>\{doc\}</document>\textbackslash n<latent\_property>\{descr\}</latent\_property>\textbackslash nPlease now generate the queries.}
\end{tcolorbox}
\end{subfigure}
\caption{Stage B descriptive lens, used for the Twitter-Conflict task. The output queries describe a latent stance without reusing topical vocabulary from the document, forcing the retriever to learn the stance axis rather than topical match.}
\label{fig:prompt_stageB_desc}
\end{figure*}

\begin{figure*}[ht]
\centering
\begin{subfigure}[t]{\textwidth}
\footnotesize
\centering
\begin{tcolorbox}[width=\linewidth,
                  colback=blue!0!white, colframe=orange!75!black,
                  title=\sys Stage B: analogue lens (Math and Writing),
                  fonttitle=\bfseries]
\textbf{System prompt:} You are given a document and a description of its abstract STRUCTURE: the underlying reasoning technique (for math problems) or stylistic fingerprint (for prose). Your job is to write \texttt{\{num\_questions\}} search queries such that any OTHER document sharing this same abstract structure (regardless of surface topic) is relevant.

\vspace{0.5em}
Each query MUST:
\begin{itemize}
\itemsep0pt\parsep0pt
\item Treat the given document itself as the example; phrase the query as ``Given this example: <short paraphrase>, retrieve <something> that shares <abstract structure described>''.
\item Describe the abstract shared structure explicitly while NOT betraying the specific topic, numbers, names, or rare vocabulary.
\item Be answerable for any analogue across different topics.
\end{itemize}

After thinking, output JSON with key \texttt{hard\_query}.

\vspace{0.5em}
\textbf{User prompt:} \texttt{<document>\{doc\}</document>\textbackslash n<abstract\_structure>\{descr\}</abstract\_structure>}
\end{tcolorbox}
\end{subfigure}
\caption{Stage B analogue lens, used for the Math Meta-Program and Writing-Style tasks. The output queries phrase the source document as an example and request other documents sharing the abstract structure across different topics.}
\label{fig:prompt_stageB_analogue}
\end{figure*}

\begin{figure*}[ht]
\centering
\begin{subfigure}[t]{\textwidth}
\footnotesize
\centering
\begin{tcolorbox}[width=\linewidth,
                  colback=blue!0!white, colframe=orange!75!black,
                  title=\sys Stage B: tip-of-the-tongue lens (Congress),
                  fonttitle=\bfseries]
\textbf{System prompt:} You are given a passage and a scenario summary that describes the rhetorical dynamic of the passage. Your job is to write \texttt{\{num\_questions\}} TIP-OF-THE-TONGUE search queries: a user's fuzzy, partial, lossy recollection of the passage, such that the passage is the ONE relevant target.

\vspace{0.5em}
Each query MUST:
\begin{itemize}
\itemsep0pt\parsep0pt
\item Sound like a real user trying to recall the exchange from memory, conversational, slightly hesitant.
\item Convey the dynamic, emotional register, and rhetorical move of the passage.
\item Be vocabulary-disjoint from the passage: NO names, NO dates, NO committees, NO verbatim phrases.
\end{itemize}

After thinking, output JSON with key \texttt{hard\_query}.

\vspace{0.5em}
\textbf{User prompt:} \texttt{<passage>\{doc\}</passage>\textbackslash n<scenario\_summary>\{descr\}</scenario\_summary>}
\end{tcolorbox}
\end{subfigure}
\caption{Stage B tip-of-the-tongue lens, used for the Congress Hearings task. The output queries mimic a user's lossy recollection, forcing the retriever to match abstract rhetorical dynamics rather than verbatim phrases.}
\label{fig:prompt_stageB_tot}
\end{figure*}

\subsection{Per-task instruction prefix}
At training and inference time, each query is prepended with a task-specific instruction following the two-element query format used by ReasonIR \citep{reasonir}:
\begin{itemize}
\item Twitter: \emph{Retrieve tweets that implicitly express the latent stance described in the query, without surface mention of the topic.}
\item Math: \emph{Given an example problem, retrieve other math problems whose solutions share the same abstract reasoning technique.}
\item Writing: \emph{Given a prose snippet, retrieve other snippets by the same author across different topics, based on stylistic fingerprint.}
\item Congress: \emph{Retrieve the specific Congressional hearing passage matching the user's lossy recollection of the exchange.}
\end{itemize}

\section{Hyperparameters and training details}
\label{app:hyperparams}

\begin{table}[t]
\centering
\small
\setlength{\tabcolsep}{6pt}
\renewcommand{\arraystretch}{1.05}
\caption{Training hyperparameters for the paper-final OBLIQ-IR retriever. The training mix is fixed and contains synthetic queries from all four tasks plus 5{,}000 authorship-derived pairs for Writing.}
\label{tab:app-hyperparams}
\begin{adjustbox}{max width=0.45\textwidth}
\begin{tabular}{ll}
\toprule
\textbf{Setting} & \textbf{Value} \\
\midrule
Backbone & \xr{\texttt{nvidia/llama-nv-embed-reasoning-3b}} \\
Pooling & mean over token embeddings, then $L_2$-normalize \\
Output dim & 3072 \\
LoRA $r$ & 16 \\
LoRA $\alpha$ & 32 \\
LoRA dropout & 0.05 \\
LoRA targets & q,k,v,o,gate,up,down projections \\
Trainable params & 24M \\
\xr{Sequence length (training, q \& d)} & \xr{1024 tokens} \\
\xr{Query length (evaluation encoding)} & \xr{256 tokens} \\
Loss & CachedMultipleNegativesRanking \\
Temperature $\tau$ & \xr{$1/0.02$} \\
Negatives per anchor & 4 mined + in-batch \\
Optimizer & AdamW, $\beta=(0.9,0.999)$ \\
Learning rate & $2\times10^{-4}$ \xr{(peak, linear decay)} \\
Warmup ratio & 0.03 \\
Batch / device & 4 \\
GPUs & 40 H100 (10 nodes $\times$ 4 GPUs) \\
Global batch & \xr{320} anchors \\
Epochs & 1 \\
Total training rows & 154{,}361 \\
Training wall-clock & approx.\ 50 min \\
Mixed precision & bf16 \\
\bottomrule
\end{tabular}
\end{adjustbox}
\end{table}

\section{Bootstrap confidence intervals}
\label{app:bootstrap}

We compute paired bootstrap 95\% confidence intervals over per-query NDCG@10 with 10000 resamples. Two contrasts are reported per task. \textbf{Recipe contribution} = (full pipeline with $\pi_t^\ast$) $-$ (no-distillation ablation, removing $\mathcal{D}_{\mathrm{wr}}^{\,\phi}$). \textbf{Reranker contribution} = (full pipeline with $\pi_t^\ast$) $-$ (dense only).

\begin{table}[t]
\centering
\small
\setlength{\tabcolsep}{8pt}
\renewcommand{\arraystretch}{1.05}
\caption{Paired bootstrap 95\% confidence intervals on per-query NDCG@10 deltas. The recipe contribution is significant on Writing and Congress and marginal on Math and Twitter; the reranker contribution is significant on all three tasks where reranking is applied. Writing's reranker contribution is zero by policy choice (identity).}
\label{tab:app-bootstrap}
\begin{adjustbox}{max width=0.45\textwidth}
\begin{tabular}{lrrrr}
\toprule
Task & $\Delta$ & 95\% CI & $p$ & $n$ \\
\midrule
\multicolumn{5}{l}{\textit{Recipe contribution (full vs no distillation)}} \\
Math & +0.024 & $[-0.003, +0.050]$ & 0.084 & 151 \\
Twitter & +0.020 & $[-0.001, +0.040]$ & 0.056 & 281 \\
Writing & +0.116 & \xr{$[+0.103, +0.129]$} & $<$0.001 & 512 \\
Congress & +0.085 & \xr{$[+0.049, +0.121]$} & $<$0.001 & 254 \\
\midrule
\multicolumn{5}{l}{\textit{Reranker contribution (full vs dense only)}} \\
Math & +0.031 & \xr{$[+0.008, +0.055]$} & 0.007 & 151 \\
Twitter & +0.027 & \xr{$[+0.006, +0.047]$} & 0.014 & 281 \\
Writing & 0 (identity) & --- & --- & 512 \\
Congress & +0.094 & \xr{$[+0.060, +0.128]$} & $<$0.001 & 254 \\
\bottomrule
\end{tabular}
\end{adjustbox}
\end{table}

\section{LLM reranking is task-dependent and non-monotonic}
\label{app:rerank-cost-quality}

\begin{figure*}[t]
\centering
\includegraphics[width=0.9\textwidth]{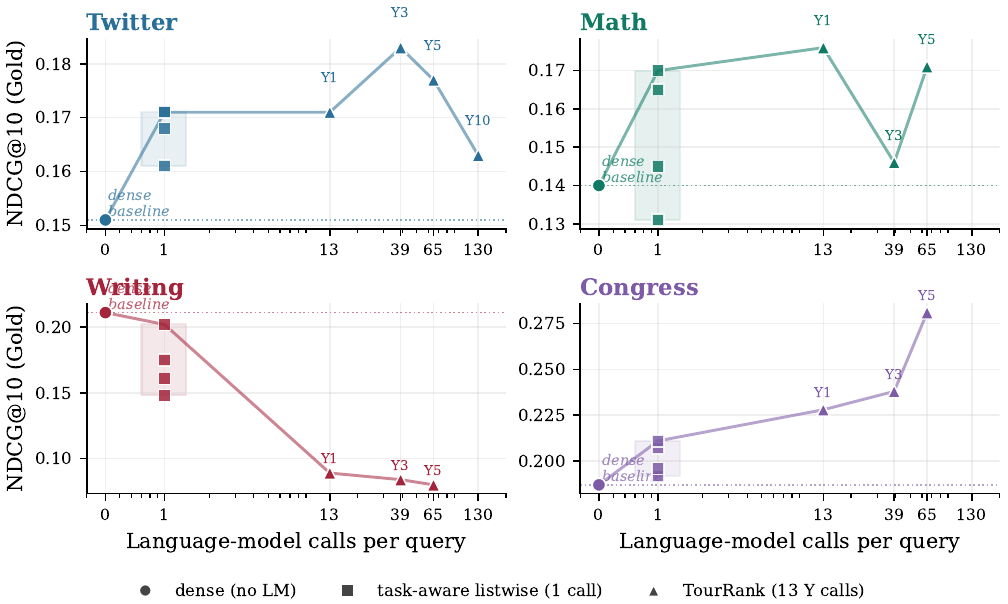}
\caption{\textbf{Cost-quality tradeoff per OBLIQ task.} NDCG@10 Gold as a function of language-model calls per query, for OBLIQ-IR dense (no reranker, circle), task-aware listwise (one call, square), and TourRank (varying $Y$, triangle). The dashed line in each panel is the dense baseline. Writing declines monotonically with reranker compute; Math and Twitter are non-monotonic in TourRank depth with peaks below the paper default $Y{=}5$; Congress increases monotonically.}
\label{fig:cost-quality}
\end{figure*}

The per-task reranker policy in Table~\ref{tab:rerank-policies} hides a richer story: even within a single reranker family, the optimal compute budget is task-specific. Figure~\ref{fig:cost-quality} sweeps both reranker families and shows NDCG@10 as a function of the number of language-model calls per query for each task. Three patterns emerge. \textbf{Writing} declines monotonically with reranker compute: more rounds of TourRank push the ranking further away from style matches and towards topic matches. \textbf{Math} and \textbf{Twitter} show a clear non-monotonic curve under TourRank, with Math peaking at $Y{=}1$ (NDCG@10 0.176, above our paper-final $Y{=}5$ value of 0.171) and Twitter peaking at $Y{=}3$ (0.183, declining to 0.163 at $Y{=}10$). \textbf{Congress} alone is monotonically increasing in $Y$ within the range we tested. The implication is that a task-uniform $Y$ leaves performance on the table on every task except Congress; a per-task $Y$ chosen on a development set is essentially free compute optimisation.

\section{Listwise reranker depth sweep}
\label{app:listwise-k}

We sweep $K$, the number of dense top-$K$ candidates passed to the task-aware listwise reranker in a single language-model call. NDCG@10 Gold.

\begin{table}[t]
\centering
\small
\setlength{\tabcolsep}{10pt}
\renewcommand{\arraystretch}{1.05}
\caption{Task-aware listwise reranker NDCG@10 Gold as a function of input depth $K$. The optimum is task-specific: Math and Twitter peak at $K{=}20$, Writing prefers $K{=}10$ (the smaller candidate set forces fine-grained stylistic attention), and Congress peaks at $K{=}50$. $K{=}100$ saturates the language model on every task except Twitter.}
\label{tab:app-listwise-k}
\begin{adjustbox}{max width=0.45\textwidth}
\begin{tabular}{ccccc}
\toprule
\textbf{K} & \textbf{Twitter} & \textbf{Math} & \textbf{Writing} & \textbf{Congress} \\
\midrule
10  & .161 & .145 & \textbf{.202} & .192 \\
20  & \textbf{.171} & \textbf{.170} & .175 & .207 \\
50  & .168 & .165 & \xr{.160} & \textbf{.211} \\
100 & .171 & .131 & .148 & .196 \\
\bottomrule
\end{tabular}
\end{adjustbox}
\end{table}

\section{TourRank tournament-depth sweep}
\label{app:tourrank-y}

We sweep $Y$, the number of TourRank tournaments per query (each tournament makes 13 language-model calls), for the four tasks. NDCG@10 Gold.

\begin{table}[ht]
\centering
\small
\setlength{\tabcolsep}{10pt}
\renewcommand{\arraystretch}{1.05}
\caption{TourRank NDCG@10 Gold as a function of tournament depth $Y$. TourRank is non-monotonic in $Y$ on three of the four tasks, with task-specific peaks: Math at $Y{=}1$, Twitter at $Y{=}3$, Congress monotonically rises through $Y{=}5$, Writing monotonically declines (every additional round pushes the ranking further from authorship and towards topic). We adopt $Y{=}5$ in the main results for comparability with the original TourRank paper.}
\label{tab:app-tourrank-y}
\begin{adjustbox}{max width=0.45\textwidth}
\begin{tabular}{lcccc}
\toprule
\textbf{$Y$ (LM calls)} & \textbf{Twitter} & \textbf{Math} & \textbf{Writing} & \textbf{Congress} \\
\midrule
1 (13)   & .171 & \textbf{.176} & .089 & .228 \\
3 (39)   & \textbf{.183} & .146 & .084 & .238 \\
5 (65)   & .177 & .171 & .080 & \textbf{.281} \\
\bottomrule
\end{tabular}
\end{adjustbox}
\end{table}

\section{\xr{Teacher-graph ablations}}
\label{app:teacher-ablations}

\xr{This section isolates what the frozen teacher's graph contributes, independently of the extra training rows it brings. Every run below uses the identical recipe, dose (5{,}000 writing pairs), anchors, negatives, instruction prefix, and effective batch size as the paper model; only the 5{,}000 distillation rows change. Because these runs were trained on one node rather than the paper's ten, we also rerun the \emph{unchanged} paper mixture under the same setup as a control: it reproduces the submitted model within 0.011 NDCG@10 on every task, which bounds run-to-run variance for the comparisons below.}

\begin{table*}[ht]
\centering
\small
\caption{Teacher-graph ablations. Dense-retriever NDCG@10 (Gold); only the 5{,}000 writing distillation rows differ between runs. The random-graph control keeps the same anchors, negatives and dose but replaces the teacher's neighbours with uniformly random corpus documents.}
\label{tab:app-teacher-ablations}
\begin{adjustbox}{max width=0.85\textwidth}
\begin{tabular}{lcccc}
\toprule
\textbf{Distillation source (5{,}000 writing rows)} & \textbf{Writing} & \textbf{Math} & \textbf{Twitter} & \textbf{Congress} \\
\midrule
Paper mixture, control rerun ($k_\phi{=}3$) & .223 & .132 & .156 & .187 \\
Random graph (random positives) & .001 & .140 & .159 & .000 \\
\midrule
$k_\phi = 1$ & .203 & .144 & .153 & .181 \\
$k_\phi = 3$ (control rerun, above) & .223 & .132 & .156 & .187 \\
$k_\phi = 5$ & .234 & .142 & .156 & .180 \\
$k_\phi = 10$ & \textbf{.249} & .142 & .154 & .179 \\
\midrule
Second teacher checkpoint of \citet{luarmud} ($k_\phi{=}3$) & .239 & .142 & .153 & .192 \\
\midrule
\textit{Reference:} paper run (\S\ref{sec:experiments}) & .211 & .140 & .151 & .187 \\
\textit{Reference:} no distillation & .096 & .148 & .158 & .196 \\
\bottomrule
\end{tabular}
\end{adjustbox}
\end{table*}

\paragraph{\xr{The graph's structure carries the signal, not the extra rows.}}
Replacing the teacher's neighbours with uniformly random writing-corpus documents, at identical anchors, negatives, dose and instruction, does not merely fail to help: it is destructive. Writing collapses from .223 to .001 and Congress from .187 to .000, while the two short-document tasks are unaffected. Inspecting the collapsed model shows textbook representational collapse: 366 of the 512 writing queries retrieve the same top-1 document, and the whole query set produces only 20 distinct top-1 documents, i.e.\ the encoder has learned that any long passage of prose matches any other. This rules out the alternative explanation that any additional writing-domain rows would help.

\paragraph{\xr{The effect is robust to the neighbourhood size.}}
At a fixed dose of 5{,}000 pairs, Writing NDCG@10 is .203 / .223 / .234 / .249 for $k_\phi = 1 / 3 / 5 / 10$, with the other three tasks stable to within 0.01 of the control. Every setting sits far above the .096 no-distillation floor, so $k_\phi$ is not a knife-edge choice; at a fixed dose, larger neighbourhoods are if anything slightly better on Writing.

\paragraph{\xr{The effect is a property of the teacher family, not of one checkpoint.}}
\citet{luarmud} release their authorship encoder in more than one checkpoint, trained on different authorship corpora. Substituting a second checkpoint at the identical dose and $k_\phi{=}3$ preserves the full effect (Writing .239, other tasks unchanged). We note also that the paper's setting is already a substantial domain transfer: the teacher is trained on Reddit authorship data, whereas the OBLIQ writing corpus is long-form prose, and the inductive bias survives that shift.

\section{\xr{Choosing a specialist: kNN-graph purity}}
\label{app:purity}

\xr{The distillation recipe requires a frozen specialist whose neighbourhood structure is aligned with the target latent attribute. This section gives a cheap test for whether a candidate specialist qualifies, \emph{before} any training is run.}

\xr{We measure the \emph{purity} of the teacher's kNN graph against gold same-author groups, recovered from the writing qrels by union--find. These groups are used for this post-hoc analysis only and never enter training. Of the teacher's 31{,}167 directed edges ($k_\phi{=}3$) over the 10{,}389-snippet corpus, 915 have both endpoints inside the labelled subset; Table~\ref{tab:app-purity} reports the result.}

\begin{table}[ht]
\centering
\small
\caption{Purity of the frozen teacher's kNN graph against gold same-author groups on the writing corpus.}
\label{tab:app-purity}
\begin{tabular}{lrr}
\toprule
\textbf{Pair source} & \textbf{Pairs} & \textbf{Same-author} \\
\midrule
Teacher kNN edges ($k_\phi{=}3$) & 915 & \textbf{45.2\%} \\
Random document pairs & 2{,}000{,}000 & 1.8\% \\
\midrule
\multicolumn{2}{l}{\textit{Lift (kNN vs.\ random)}} & \textbf{25.4}$\times$ \\
\bottomrule
\end{tabular}
\end{table}

\xr{This yields a practical selection recipe for a new latent-attribute task: encode a small labelled probe set with any candidate specialist, measure the same-attribute rate among its nearest neighbours against the random-pair base rate, and distil only if the lift is large. The test costs one encoding pass and no training. The random-graph control in Table~\ref{tab:app-teacher-ablations} validates the diagnostic end to end: the 1.8\%-purity graph collapses Writing to .001, whereas the 45.2\%-purity graph reaches .223.}

\section{\xr{Synthetic query quality}}
\label{app:query-quality}

\xr{Section~\ref{sec:permech} instructs the generator to avoid named entities and rare vocabulary drawn from the source document. This section verifies that the instruction is followed. We compute, for every synthetic training query against its source document, the content-word Jaccard overlap and the fraction of the query's content words that also appear in the document, plus an entity-leakage rate. Content words are alphabetic tokens longer than three characters with stopwords removed; entities are detected heuristically as non-sentence-initial capitalised spans plus four-digit years. As a reference point we run the identical measurement on the official OBLIQ queries against their gold documents; Table~\ref{tab:app-query-quality} reports both.}

\begin{table}[ht]
\centering
\small
\caption{Lexical overlap and entity leakage of our synthetic training queries against their source documents, with the official OBLIQ queries against their gold documents as a reference. Lower is more oblique. $^\ast$Official Writing ``queries'' are raw prose snippets (the task is query-by-example), so they necessarily contain the source's entities.}
\label{tab:app-query-quality}
\begin{adjustbox}{max width=0.45\textwidth}
\begin{tabular}{lcccccc}
\toprule
& \multicolumn{2}{c}{\textbf{Jaccard}} & \multicolumn{2}{c}{\textbf{Query-word coverage}} & \multicolumn{2}{c}{\textbf{Entity-leak rate}} \\
\cmidrule(lr){2-3} \cmidrule(lr){4-5} \cmidrule(lr){6-7}
\textbf{Task} & synthetic & official & synthetic & official & synthetic & official \\
\midrule
Twitter  & .016 & .002 & .036 & .004 & .017 & .014 \\
Math     & .070 & .066 & .097 & .081 & .057 & .091 \\
Writing  & .021 & .054 & .105 & .106 & .030 & .509$^\ast$ \\
Congress & .043 & .030 & .359 & .102 & .064 & .098 \\
\bottomrule
\end{tabular}
\end{adjustbox}
\end{table}

\xr{Entity leakage is low throughout (1.7--6.4\%), and lexical overlap is comparable to the benchmark's own oblique queries on three of the four tasks. We report one exception honestly: our synthetic Congress recollections share considerably more common vocabulary with their source passage than the officially hardened queries do (.359 vs.\ .102 coverage). Because evaluation always uses the official, vocabulary-disjoint queries and the trained model transfers to them (Congress .281 vs.\ .183 for the strongest published baseline), the model is not simply exploiting a lexical shortcut; but the gap indicates that the tip-of-the-tongue lens should be hardened further, and we flag it as the weakest point of the generation pipeline.}

\section{\xr{The frozen teacher as a direct retriever}}
\label{app:teacher-baseline}

\xr{Because the writing task is query-by-example (its queries are prose snippets), the frozen authorship encoder can be used as a retriever directly, without any training. This is the natural upper reference for what the distillation transfers, and we report it as a baseline in Table~\ref{tab:writing-results}. Encoding path, truncation, official queries and excluded-id masking are identical to our own evaluation.}

\xr{The teacher alone is a strong baseline: at .206 NDCG@10 it exceeds every published non-oracle number on Writing, including Gemini-2-Embedding at .164. The student nonetheless matches it at the top of the ranking (.211, $+$.005) and clearly surpasses it deeper down ($+$.024 NDCG@50, $+$.061 Recall@100), despite having been trained on only 5{,}000 of the teacher's 31{,}167 graph edges. Distillation therefore does not merely clone the specialist: it composes the specialist's style axis with the student's general retrieval pretraining. We stress that this baseline exists only on Writing. An authorship encoder cannot serve the descriptive, analogue or tip-of-the-tongue natural-language queries of the other three tasks, whereas OBLIQ-IR is a single shared encoder over a single index for all four.}

\section{\xr{Full benchmark results, including WildChat-Errors}}
\label{app:wildchat}

\xr{For completeness we report our numbers on the fifth OBLIQ-Bench task, WildChat-Errors, which we exclude from all claims (Limitations). Table~\ref{tab:app-wildchat} gives the full comparison.}

\begin{table}[ht]
\centering
\small
\setlength{\tabcolsep}{6pt}
\renewcommand{\arraystretch}{1.05}
\caption{WildChat-Errors, NDCG@10 Gold / Pooled. Baselines reproduced from \citet{oblique2026}. OBLIQ-IR does not improve on this task and we make no claim on it.}
\label{tab:app-wildchat}
\begin{tabular}{lc}
\toprule
\textbf{Pipeline} & \textbf{NDCG@10 (G/P)} \\
\midrule
BM25 & .004 / .006 \\
LateOn-0.1B & .002 / .003 \\
Qwen3-Embed-0.6B & .059 / .073 \\
Qwen3-Embed-4B & .031 / .059 \\
Gemini-2-Embedding & .057 / .097 \\
GPT-5.2 Query Rewriter & .065 / .095 \\
GPT-5.2 Multi-Hop Agent & \textbf{.070 / .113} \\
\midrule
OBLIQ-IR (no distillation) & .047 / .046 \\
OBLIQ-IR (paper-final) & .026 / .025 \\
\midrule
Oracle GPT-5.2 Tournament & .397 / .431 \\
\bottomrule
\end{tabular}
\end{table}

\xr{Three factors make this task different in kind from the four we report. First, the best published non-oracle system reaches only .070/.113, and \citet{oblique2026} observe that multi-hop search helps WildChat far less than the other tasks because the failure mode is distributed across many turns of a long conversation rather than localised in a passage. Second, our dense Recall@100 is .040 over a 507{,}729-document corpus, so the task is recall-bound in a way that none of our training signals addresses. Third, the split has only 40 queries, so differences of this magnitude carry little statistical power. Every training-data variant we tried moved WildChat down rather than up, and we judged it inappropriate to claim the task.}

\section{Random-shuffle sanity check}
\label{app:shuffle}

To verify that the reported NDCG gains are not artifacts of how the qrels overlap with the dense top-100, we shuffle each task's dense top-100 uniformly at random and re-evaluate. The dense-to-shuffled NDCG@10 ratio is the signal-to-noise of the dense ordering.

\begin{table}[t]
\centering
\small
\caption{Random-shuffle sanity check. Every task is well above the chance baseline; Congress and Twitter have the strongest dense-ordering signal.}
\label{tab:app-shuffle}
\begin{adjustbox}{max width=0.45\textwidth}
\begin{tabular}{lccr}
\toprule
\textbf{Task} & Dense NDCG@10 & Random-shuffle NDCG@10 & Ratio \\
\midrule
Twitter  & .151 & .024 & 6.3$\times$ \\
Math     & .140 & .057 & 2.5$\times$ \\
Writing  & .211 & .043 & 4.9$\times$ \\
Congress & .187 & .020 & 9.4$\times$ \\
\bottomrule
\end{tabular}
\end{adjustbox}
\end{table}

\section{What we tried that did not work}
\label{app:negatives}

We briefly list experiments that did not improve over OBLIQ-IR, in case they save other researchers time.
\begin{enumerate}
\item \xr{Increasing the distillation dose from 5{,}000 pairs to the full 31{,}167-edge kNN graph overfits Writing while collapsing Congress; the per-task numbers are in Table~\ref{tab:app-dose} and Figure~\ref{fig:method} (right).}
\item Pure MarginMSE distillation from a cross-encoder reranker collapses every task by 0.07 to 0.20 NDCG@10 because the in-batch contrastive signal is lost.
\item A hybrid InfoNCE + MarginMSE loss restores Math and Twitter but does not improve Writing and slightly hurts Congress.
\item Longer-context training (2048 tokens versus 1024) does not change any task by more than 0.005.
\item A task-conditional listwise reranker trained with knowledge distillation underperforms the inference-only TourRank on all three topical tasks.
\item A stricter task-aware listwise prompt for Writing that explicitly enumerates the style features to attend to (sentence rhythm, register, punctuation tics) and the features to ignore (subject, plot, entities) reaches 0.158 NDCG@10, \emph{worse} than the original task-aware listwise prompt (0.175) and well below dense-only (0.211); explicit prompting cannot escape the language model's topical bias.
\item A cascade reranker that applies TourRank to the dense top-$200$ instead of top-$100$ hurts Math by $-0.010$ NDCG@10 and ties Twitter; the larger candidate set dilutes the per-document share of the tournament budget.
\end{enumerate}

\begin{table}[t]
\centering
\small
\caption{\xr{Distillation dose sweep, dense-retriever NDCG@10 (Gold). ``Blend'' adds the full teacher graph to the existing synthetic writing rows; ``replace'' substitutes it for them. The 5{,}000-pair dose used by OBLIQ-IR is the only setting in which no task degrades materially: larger doses buy Writing at the cost of Congress.}}
\label{tab:app-dose}
\begin{adjustbox}{max width=0.45\textwidth}
\begin{tabular}{lcccc}
\toprule
\textbf{Distillation dose (writing pairs)} & \textbf{Writing} & \textbf{Math} & \textbf{Twitter} & \textbf{Congress} \\
\midrule
0 (no distillation) & .096 & \textbf{.148} & \textbf{.158} & \textbf{.196} \\
5{,}000 (paper-final) & .211 & .140 & .151 & .187 \\
31{,}167 added (blend) & .331 & .134 & .149 & .091 \\
31{,}167 replacing (replace) & \textbf{.374} & .144 & .154 & .055 \\
\bottomrule
\end{tabular}
\end{adjustbox}
\end{table}

\section{Distance to the oracle ceiling}
\label{app:ceiling}

The Oracle GPT-5.2 Tournament \citep{oblique2026}, a verification-only oracle that has access to all gold judgments at scoring time, reaches 0.279 NDCG@10 on Math, 0.331 on Twitter, 0.515 on Writing, and 0.913 on Congress. The remaining gap between OBLIQ-IR (with the chosen per-task reranker) and the oracle is therefore approximately 0.108 (Math), 0.154 (Twitter), 0.304 (Writing), and 0.632 (Congress). Most of the gap is on Congress, where the dense Recall@100 of 0.311 already bounds any rerank-based improvement; this is the same recall-ceiling phenomenon as Writing (Section~\ref{sec:recall-ceiling}) but at a much lower starting point because the corpus is small and the queries are very lossy. Closing this gap will require either retrieval-time language-model use (which the agentic baselines already attempt) or a different first-stage architecture.

\section{Reranker prompts}
\label{app:rerank-prompts}

This section gives the prompts behind the per-task reranker policies of
Table~\ref{tab:rerank-policies}. Two families are used. The \emph{generic
tournament} prompt (Figure~\ref{fig:prompt_tourrank}) is the unmodified prompt of
\citet{tourrank}: it names no latent attribute and delegates aggregation to the
tournament structure. It is the reranker adopted for Math, Twitter, and Congress.
The \emph{task-aware listwise} prompts
(Figures~\ref{fig:prompt_listwise_math}--\ref{fig:prompt_listwise_congress}) are
ours: one per task, each naming the latent attribute $f_t$ explicitly and
instructing the model against a topical fallback. Writing uses the identity policy
in the paper-final pipeline, so Figure~\ref{fig:prompt_listwise_writing} is
reported for completeness and for the ablation of
Appendix~\ref{app:negatives} rather than as part of the final system.

\begin{figure*}[ht]
\centering
\begin{subfigure}[t]{\textwidth}
\footnotesize
\centering
\begin{tcolorbox}[width=\linewidth,
                  colback=blue!0!white, colframe=gray!70!black,
                  title=\sys generic TourRank tournament prompt,
                  fonttitle=\bfseries]
\textbf{System prompt:} You are an expert ranker. You will receive a query and a list of candidate documents. Read each candidate and pick the top \texttt{\{M\}} that are most relevant to the query. Return only the document IDs.

\vspace{0.5em}
\textbf{Settings.} $Y{=}5$ shuffled tournaments per query; each tournament makes 13 LM calls; input pool is the dense top-100; output is the top-10.
\end{tcolorbox}
\end{subfigure}
\caption{TourRank's per-call prompt \citep{tourrank}. The prompt is intentionally generic; the tournament structure handles aggregation. This is the reranker used for Math, Twitter, and Congress in the paper-final pipeline.}
\label{fig:prompt_tourrank}
\end{figure*}

\begin{figure*}[ht]
\centering
\begin{subfigure}[t]{\textwidth}
\footnotesize
\centering
\begin{tcolorbox}[width=\linewidth,
                  colback=blue!0!white, colframe=teal!70!black,
                  title=\sys task-aware listwise reranker: Math,
                  fonttitle=\bfseries]
\textbf{System prompt:} You are ranking math documents for ANALOGOUS PROBLEM-SOLVING TECHNIQUE. The query describes a math problem and its solution approach.

\vspace{0.5em}
Pick documents that use the SAME mathematical reasoning method, the same transformation, the same key insight, NOT documents about the same topic.

\vspace{0.5em}
\textbf{User prompt:} Query, list of 20 candidates; return the ranking as a comma-separated list of IDs.
\end{tcolorbox}
\end{subfigure}
\caption{Task-aware listwise reranker for Math. The prompt names the latent attribute explicitly (proof technique) and tells the LM not to fall back on topical similarity.}
\label{fig:prompt_listwise_math}
\end{figure*}

\begin{figure*}[ht]
\centering
\begin{subfigure}[t]{\textwidth}
\footnotesize
\centering
\begin{tcolorbox}[width=\linewidth,
                  colback=blue!0!white, colframe=blue!60!black,
                  title=\sys task-aware listwise reranker: Twitter,
                  fonttitle=\bfseries]
\textbf{System prompt:} You are ranking tweets for SAME IMPLICIT STANCE TOWARD A GEOPOLITICAL CONFLICT. The query expresses a stance about the conflict implicitly: through framing, irony, evaluative attitude, or rhetorical pose, not through explicit pro/against statements.

\vspace{0.5em}
Pick tweets that express the SAME implicit stance toward the same conflict, even if the surface vocabulary differs. A tweet that overlaps with the query on entities or events but expresses a different (or no) stance is NOT relevant. Sarcasm and irony should be matched on actual stance, not on surface text.

\vspace{0.5em}
\textbf{User prompt:} Query, list of 20 candidates; return the ranking as a comma-separated list of IDs.
\end{tcolorbox}
\end{subfigure}
\caption{Task-aware listwise reranker for Twitter. The prompt frames relevance as opinion alignment rather than topical overlap.}
\label{fig:prompt_listwise_twitter}
\end{figure*}

\begin{figure*}[ht]
\centering
\begin{subfigure}[t]{\textwidth}
\footnotesize
\centering
\begin{tcolorbox}[width=\linewidth,
                  colback=blue!0!white, colframe=red!70!black,
                  title=\sys task-aware listwise reranker: Writing,
                  fonttitle=\bfseries]
\textbf{System prompt:} You are ranking literary documents for SAME AUTHOR / SAME WRITING STYLE. The query is a passage.

\vspace{0.5em}
Pick documents that share the SAME AUTHORIAL STYLE (vocabulary register, sentence rhythm, punctuation habits, characteristic word choice) as the query, EVEN IF on a completely different topic. IGNORE topical similarity, only style/voice matters.

\vspace{0.5em}
\textbf{User prompt:} Query, list of 20 candidates; return the ranking as a comma-separated list of IDs.
\end{tcolorbox}
\end{subfigure}
\caption{Task-aware listwise reranker for Writing. The prompt is the most explicit of the four: it enumerates the style features to attend to (rhythm, register, punctuation, lexicon) and forbids any topical reasoning. Even with this instruction the reranker fails to match the dense-only baseline on Writing (see Table~\ref{tab:rerank-policies} and Appendix~\ref{app:negatives}).}
\label{fig:prompt_listwise_writing}
\end{figure*}

\begin{figure*}[ht]
\centering
\begin{subfigure}[t]{\textwidth}
\footnotesize
\centering
\begin{tcolorbox}[width=\linewidth,
                  colback=blue!0!white, colframe=purple!70!black,
                  title=\sys task-aware listwise reranker: Congress,
                  fonttitle=\bfseries]
\textbf{System prompt:} You are ranking Congressional speeches for TIP-OF-TONGUE RETRIEVAL. The query is a partial, imprecise, or fragmented description of a speech.

\vspace{0.5em}
Pick speeches that match the LATENT DESCRIPTION even if the surface wording differs. The user remembers some features (themes, time period, speaker traits) but not exact phrases.

\vspace{0.5em}
\textbf{User prompt:} Query, list of 20 candidates; return the ranking as a comma-separated list of IDs.
\end{tcolorbox}
\end{subfigure}
\caption{Task-aware listwise reranker for Congress. The prompt instructs the LM to match latent rhetorical descriptions against speeches whose surface wording is different.}
\label{fig:prompt_listwise_congress}
\end{figure*}

\end{document}